\documentclass[preprint2]{aastex}
\pdfoutput=1

\usepackage{amsmath,amssymb,amsfonts} 
\usepackage{color} 
\usepackage{hyperref}
\hypersetup{
linktocpage=true,
pdfborderstyle={/S/S/W 1},
hyperindex=true,
bookmarks=true,
bookmarksopen=true,
bookmarksnumbered=true}
\usepackage{epsfig}
\usepackage{natbib}

\newcommand \beq {\begin{equation}}
\newcommand \enq {\end{equation}}

\newcommand \qloss {\mathcal{L}}
\newcommand \tcool {t_{\rm {cool}}}
\newcommand \tceff {\langle \langle t_{\rm {cool}} \rangle \rangle_t}
\newcommand \aone  {\langle\alpha\rangle_{t}}

\newcommand \reynolds {\rho v_{x} \delta v_{y}}
\newcommand \newton {g_{x} g_{y}/4\pi G}

\slugcomment{}

\shorttitle{Gravito-Turbulence in 3D}
\shortauthors{Shi \& Chiang}

\begin{document}


\title{Gravito-Turbulent Disks in 3D: Turbulent Velocities vs.~Depth}


\author{Ji-Ming Shi\altaffilmark{1,2} and Eugene Chiang\altaffilmark{1,2,3}}
\altaffiltext{1}{Department of Astronomy, UC Berkeley, Hearst Field Annex B-20,
    Berkeley, CA 94720-3411}
\altaffiltext{2}{Center for Integrative Planetary Science, UC Berkeley, Hearst Field Annex B-20,
    Berkeley, CA 94720-3411}
\altaffiltext{3}{Department of Earth and Planetary Science, UC Berkeley, 307 McCone Hall,
    Berkeley, CA 94720-4767}
    
\email{jmshi@berkeley.edu}

\begin{abstract}
  Characterizing turbulence in protoplanetary disks is crucial for
  understanding how they accrete and spawn planets.  Recent
  measurements of spectral line broadening promise to diagnose
  turbulence, with different lines probing different depths. We use 3D
  local hydrodynamic simulations of cooling, self-gravitating disks to
  resolve how motions driven by ``gravito-turbulence'' vary with
  height.  We find that gravito-turbulence is practically as vigorous
  at altitude as at depth. Even though gas at altitude is much too
  rarefied to be itself self-gravitating, it is strongly forced by
  self-gravitating overdensities at the midplane.  The long-range
  nature of gravity means that turbulent velocities are nearly uniform
  vertically, increasing by just a factor of 2 from midplane to
  surface, even as the density ranges over nearly three orders of
  magnitude.  The insensitivity of gravito-turbulence to height
  contrasts with the behavior of disks afflicted by the
  magnetorotational instability (MRI); in the latter case,
  non-circular velocities increase by at least a factor of 15 from
  midplane to surface, with various non-ideal effects only magnifying
  this factor.  The distinct vertical profiles of
  gravito-turbulence vs.~MRI turbulence may be used in conjunction
  with measurements of non-thermal linewidths at various depths to
  identify the source of transport in protoplanetary disks.
\end{abstract}



\section{INTRODUCTION \label{sec:intr}}

How protoplanetary disks transport angular momentum and mass has been 
a longstanding mystery (e.g., \citealt{hartmann06}). Turbulence driven 
by the magnetorotational instability \citep[MRI;][]{BH91,BH98} is 
perhaps the most intensively studied mechanism.  Recent studies explore a
host of non-ideal magnetohydrodynamic effects that strongly affect
the character of transport in poorly ionized disks
\citep[e.g.,][]{perezbecker11a,perezbeckerchiang11,bai11,WS2012,BS2013,Bai13,%
Simon13b,Simon13a,kunz13,lesur14}.
Disk self-gravity is another option for sufficiently massive disks
\citep[e.g.,][]{Paczynski78,gammie01,Forgan2012}.  Torques could be
exerted either by globally coherent spiral structure, or by local
density waves that are continuously generated and dissipated in
a state of ``gravito-turbulence''.
Non-circular motions, turbulent or otherwise, impact planet and
planetesimal formation insofar as they regulate grain growth
\citep[e.g.,][]{ormeletal07}, the degree to which dust settles and concentrates
\citep[e.g.,][]{leeetal10,leeetal10b}, and how planets migrate
\citep[e.g.,][]{nelsonpapa04,laughlinetal04,rein12}. 
Pinning down the nature of turbulence 
in protoplanetary disks is a first-rank problem.

Spectral line broadening offers empirical constraints on non-circular motions.
Using Submillimeter Array observations of the CO~$(3-2)$
emission line, \citet{Hughes2011} found that the ``turbulent''
linewidth (i.e., the Doppler ``$b$''-parameter\footnote{Not to be confused
with the cooling parameter $b$ introduced later in our paper.})
in the high-altitude
outskirts of the HD 163296 disk is $\sim$300 m/s, about 40\% of the
local sound speed.
This non-thermal linewidth appears consistent with MRI turbulence, but
other drivers have not been ruled out.
Lines from other transitions would enable us to plumb the depths of
turbulence vertically and thereby constrain its origin
(\citealt{Simon2011}; Hughes et al.~2014, in preparation). Similarly
illuminating would be radial profiles of non-thermal linewidths
\citep{Forgan2012}; these
are promised by, e.g., the Atacama Large Millimeter Array.

In this paper we numerically simulate gravito-turbulence in a 3D
shearing box. Our goal is to measure how the non-circular motions
generated by gravito-turbulence vary with disk altitude. By
determining what is distinctive about these vertical profiles, we hope
to inform observations of non-thermal linewidths like those pioneered
by \citet{Hughes2011}, and ultimately to characterize 
turbulence in disks, protoplanetary or otherwise.  
We utilize a grid-based code (\texttt{Athena}) to resolve
stratified, self-gravitating, secularly cooling disks, achieving
unprecedentedly high resolution and dynamic range in the vertical direction
(cf.~\citealt{Forgan2012}). Radiative cooling is treated in the
optically thin limit in which every grid cell cools independently of
every other.  We experiment with two cooling prescriptions: either the
cooling time is fixed in space and time (``constant cooling time'') or
it depends on the local temperature (``optically-thin thermal
cooling'').  Our disks attain a state of steady gravito-turbulence in
which the imposed cooling is balanced by compressive heating driven by
gravitational instability.

We describe our numerical methods in \S\ref{sec:method}. 
Results are presented in \S\ref{sec:result}, and placed into
physical context in \S\ref{sec:conclusion}, together with an outlook.

\section{METHODS\label{sec:method}}
The equations we solve are listed in \S\ref{sec:equation}; a
description of our code and our adopted boundary conditions are given
in \S\ref{sec:numerical}; initial conditions and simulation parameters
are provided in \S\ref{sec:initial}; and some averages used to
diagnose our results are defined in \S\ref{sec:diag}.

\subsection{Equations Solved\label{sec:equation}}
We solve the hydrodynamic equations governing three-dimensional,
self-gravitating, stratified accretion disks, including the effects of
secular cooling. The disk is modeled in the local shearing box
approximation. In a Cartesian reference frame corotating with the disk at fixed
orbital frequency $\Omega \hat{\mathbf{z}}$, the equations solved by
our code are as follows:
\begin{align}
  \frac{\partial \rho}{\partial t} + \nabla\cdot (\rho \mathbf{v}) = 0 \,, 
  \label{eq:continuity} \\
  \frac{\partial \rho\mathbf{v}}{\partial t} + \nabla\cdot\left(\rho\mathbf{v}\mathbf{v}
  +P\mathbf{I} + \mathbf{T_{\rm g}} \right) = 
  - 2\rho \Omega \hat{\mathbf{z}} \times\mathbf{v} \nonumber \\ 
  +2q\rho\Omega^2 x \hat{\mathbf{x}} -\rho\Omega^2 z\hat{\mathbf{z}}\,,
  \label{eq:eom} \\
  \frac{\partial E}{\partial t} + \nabla\cdot (E+P)\mathbf{v} =
  -\rho\mathbf{v}\cdot\nabla\Phi \nonumber \\
  +\rho\Omega^2\mathbf{v}\cdot\left(2 q x \hat{\mathbf{x}} - z\hat{\mathbf{z}}\right) -\rho\qloss \,,
  \label{eq:eoe} \\
  \nabla^2\Phi = 4\pi G \rho\,, 
  \label{eq:poisson}
\end{align}
where $\hat{\mathbf{x}}$ points in the radial direction,
$\rho$ is the gas mass density, $\mathbf{v}$ is the gas velocity,
$P$ is the gas pressure, $\Phi$ is the disk's self-gravitational potential,
$q = 3/2$ is the Keplerian shear parameter, 
\beq
  E = {U} + {K} = \frac{P}{\gamma -1} + \frac{1}{2}\rho v^2
  \label{eq:eos}
\enq
is the sum of the internal energy density $U$ and bulk kinetic energy
density $K$ for an ideal gas with specific heat ratio $\gamma = 5/3$, and
\beq
  \mathbf{T_{\rm g}} = \frac{1}{4\pi G}\left[\nabla\Phi\nabla\Phi -
  \frac{1}{2}\left(\nabla\Phi\right)\cdot\left(\nabla\Phi\right)\mathbf{I}\right] 
  \label{eq:grav_tensor}
\enq
is the gravitational stress tensor with identity tensor $\mathbf{I}$.

We consider two prescriptions
for the volumetric cooling rate (a.k.a.~volume emissivity)
$\rho\qloss(\rho,U)$. In the first case, we have
\beq
\rho\qloss = U/\tcool = \Omega U / \beta \,\,\,\,
{\rm (constant \,\, cooling \,\, time)}
\enq
with $\beta \equiv \Omega \tcool$ constant everywhere.
The assumption of constant cooling time $\tcool$ is adopted
by many 2D
\citep[e.g.,][]{gammie01,JG2003,Paardekooper2012} and 3D
\citep[e.g.,][]{Rice2003,LR2004,LR2005,Mejia2005,Cossins2009,MB2011}
simulations of self-gravitating disks.
This prescription enables direct experimental control over the
rate of energy loss.

In our second treatment of cooling, we assume that the cooling radiation is
thermal and that the disk is optically thin to such radiation. Then
every parcel of gas has its own cooling time $\tcool = U / (\rho \qloss) \propto T / \qloss \propto 1/(T^3\kappa)$, with $\qloss \propto T^4 \kappa$ for temperature $T$ 
and opacity $\kappa$. 
We assume constant $\kappa$, as would be the case if the 
cooling radiation were emitted by grains in the 
geometric optics limit (so that $\kappa$ is independent
of $T$), and if the grains were uniformly mixed with gas (so that $\kappa$
is independent of $\rho$). These assumptions may hold
in the outermost portions of gravito-turbulent disks, 
where opacities are dominated by mm--cm sized dust particles
and where strong vertical flows in dense gas can keep such particles aloft.
Other opacity
laws characteristic of molecules or H$^-$ may obtain in the hotter inner
regions where dust sublimates (cf.~\citealt{BL1994}).
For constant $\kappa$ we have $\tcool = b (\rho/P)^3$ for constant $b$,
or equivalently
\beq
\rho \qloss = \frac{1}{b(\gamma-1)} \frac{P^4}{\rho^3} \,\,\,\, {\rm (optically \,\, thin, \,\, thermal \,\, cooling)}.
\label{eq:qloss}
\enq
In our experiments, we choose $b$ such that the cooling time
averaged over our simulation domain equals some desired value.

\subsection{Code Description and Boundary Conditions\label{sec:numerical}}

Our simulations are run with \texttt{Athena} \citep{sg10}.  We adopt
the van Leer integrator \citep{vl06,sg09}, a piecewise linear spatial
reconstruction in the primitive variables, and the HLLC
(Harten-Lax-van Leer-Contact) Riemann solver.  We solve Poisson's
equation using fast Fourier transforms \citep{ko09,kko11}.
Self-gravity is added to the momentum equation in a conservative
form (as shown in equation~\ref{eq:eom}),
but added to the energy equation (\ref{eq:eoe}) as a source term;
thus the total energy does not conserve to round-off error
(for a more accurate algorithm, see \citealt{Jiang2013}).
We have verified, however, that the error introduced in our
    treatment of energy is negligible, as we reproduce well
    the analytic result of \citet{gammie01} for how the stress
    varies with cooling---this dependence
    essentially reflects energy conservation (see our
    \S\ref{sec:alpha_tc} and Figure~\ref{fig:a_tc_gammie}).

Boundary conditions for our hydrodynamic flow variables ($\rho$,
$\mathbf{v}$, $U$, but not the self-gravitational potential $\Phi$)
are shearing-periodic in radius ($x$), periodic in azimuth ($y$), and
outflow in height ($z$). The Poisson solver for $\Phi$ implements
shearing-periodic boundary conditions in $x$, periodic boundary
conditions in $y$, and vacuum boundary conditions in $z$
\citep{ko09,kko11}. The ghost-cell values for $\Phi$ in $z$ are set by
solving the finite-difference form of the Poisson equation. Note that
the version of \texttt{Athena}'s shearing-box Poisson solver that we
downloaded produces velocities in the simulation domain that are
discontinuous with those in ghost cells if the box is too large in the
$x$-direction and if the Courant number (governing our timestep)
$\gtrsim$ 0.4--0.5. We found that we could eliminate these
discontinuities by reducing the Courant number to $\sim$0.1, but did
not implement a deeper fix.

We use orbital advection algorithms to shorten the timestep
and improve conservation \citep{Masset2000, Johnson2008, sg10}.  Upon
adding cooling as an explicit source term in equation (\ref{eq:eoe}), we
also modify the timestep $\Delta t$ to equal $\min (\Delta t_0,
\epsilon\,\tcool)$, where $\Delta t_0$ is \texttt{Athena}'s usual
Courant-limited timestep, $\tcool$ is evaluated for every cell,
and $\epsilon=0.02$, small enough to 
resolve the cooling history.
Typically $\Delta t_0$ is $10^2$--$10^4$ times shorter than $\min\,(\tcool)$.

\subsection{Initial Conditions, Run Parameters, and Box Sizes\label{sec:initial}}
For our constant cooling simulations, we take $\beta \equiv \Omega \tcool \in \{3, 4, 5, 8,10, 20, 40, 80\}$. Initial conditions for $\beta \in \{20, 40, 80 \}$ 
are derived from $\beta = 10$ by ``morphing'': we initialize
$\beta = 20$ with the final gravito-turbulent outcome from $\beta =
10$; $\beta = 40$ is initialized with the final outcome from $\beta =
20$; and $\beta = 80$ is initialized with the final outcome from
$\beta = 40$. 
Initial conditions for $\beta < 10$ are derived directly from the 
final outcome of $\beta = 10$.
Morphing has the advantage that for each $\beta$, the disk can adjust
more quickly to its quasi-equilibrium state, i.e., we bypass as much as possible
initial violent transients. In any case,
we take care to run every simulation for long enough duration
(typically several cooling times) that a steady gravito-turbulent state is reached in which time-averaged
quantities such as rms density and velocity fluctuations have converged
to unique values.
All simulations with optically thin cooling are
initialized with the outcome from $\beta=10$.  See
Table~\ref{tab:tab1} for run parameters.

\begin{figure}[h!]
  \epsscale{1.0} \plotone{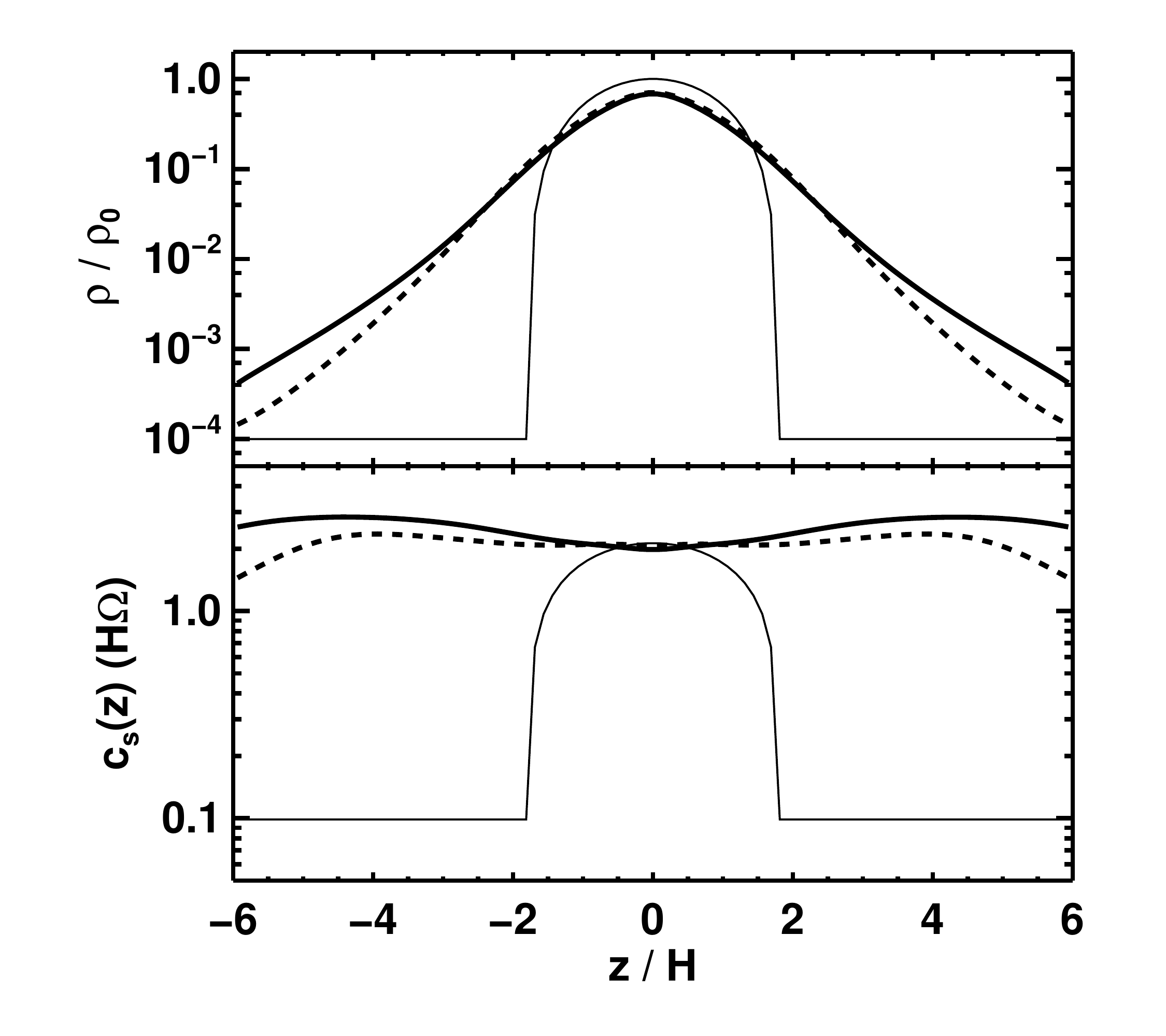} \caption{\small{\textsl{Top}:
      Density as a function of height, drawn from our static initial
      conditions (thin solid line); from the fully gravito-turbulent
      state to which our $\beta =\Omega\tcool = 10$ simulation
      (tc=10.hi) relaxes, time-averaged from $t=200$--$300\Omega^{-1}$
      (thick solid line); and from the gravito-turbulent outcome of
      our optically-thin thermal cooling simulation b=500.hi,
      similarly time-averaged (dashed line).  \textsl{Bottom:} Same as
      top but for sound speed. In both our cooling prescriptions,
      our disks are optically thin in the sense that every grid cell
      cools independently of every other grid cell; we find that such
      disks, when heated by gravito-turbulence, are nearly isothermal.}}
\label{fig:initial}
\end{figure}

For $\beta = 10$, the initial vertical density profile is derived
semi-analytically.  We solve for vertical hydrostatic
equilibrium including self-gravity (cf.~\citealt{sc2013}):
\beq
\frac{1}{\rho}\frac{dP}{dz} = -\Omega^2 z - 4\pi
G\int^z_0\rho(z^\prime)dz^\prime 
\label{eq:static1}
\enq 
for a polytropic gas $P=K \rho^{\gamma}$ (the polytropic assumption
is used only for this initialization; the subsequent evolution obeys
the full energy equation as described in \S\ref{sec:equation}).
A non-dimensional, differential form of
(\ref{eq:static1}) reads:
\beq
\frac{d^2\tilde\rho}{d\tilde z^2} + \frac{\gamma -
2}{\tilde\rho}\left(\frac{d\tilde\rho}{d\tilde z}  \right)^2 + Q_0^2\tilde{\rho}^{2-\gamma} +
\frac{2}{h}\tilde{\rho}^{3-\gamma} = 0 
\label{eq:static2}
\enq where $\tilde\rho \equiv \rho/\rho_0$, $\rho_0 \equiv \rho(z=0)$,
$\tilde z \equiv z/H_{\rm sg}$, $H_{\rm sg} \equiv [c_{\rm s}(0)]^2/(\pi G
\Sigma_0)$ is a fiducial lengthscale for a self-gravitating disk,
$c_{\rm s}(0) = \sqrt{\gamma P(z=0)/\rho_0} =
\sqrt{K\gamma\rho_0^{\gamma-1}}$ is the initial sound speed at the midplane,
$\Sigma_0 \equiv 2\rho_0 H$ is the surface density for an
effective half-thickness $H$, $h \equiv H/H_{\rm sg}$, and Toomre's
initial $Q_0 \equiv
c_{\rm s}(0)\Omega/(\pi G \Sigma_0)$. From these definitions,
\beq
h = \frac{1}{2}\int^{\infty}_{-\infty} \tilde\rho\, d\tilde{z} \,.
\label{eq:hscale}
\enq
Upon setting $Q_0=1$ and $\gamma=5/3$, we
iteratively solve equations (\ref{eq:static2}) and (\ref{eq:hscale})
for $\tilde\rho(\tilde{z})$ and $h$ (starting with an initial guess for
$h$). We find that $h = 0.4703$, and show the
density profile in Figure~\ref{fig:initial}.
From these parameters, plus our code units
($\rho_0= H = \Omega=1$),
it follows that $c_{\rm s}(0) = 2.126$
(equivalently $K = 2.712$) and $G = 0.3384$.

We verified by direct simulation that these initial conditions yield
disks that are stable to perturbations in the absence of cooling
($\beta = \infty$).
In our experiments with $\beta=10$, we initialize the disk with
random, cell-to-cell velocity perturbations up to $0.1 c_{\rm s}(0)$
for $|z| < 2H$.  These random velocities are introduced only at $t=0$;
they are not driven.  Our experiments with other $\beta$'s
(initialized according to our ``morphing'' procedure) and optically-thin
cooling (initialized with the end state of our $\beta=10$ run) begin
turbulent, and so for these
simulations we introduce no further perturbation.

Our boxes span $[-32H, 32H]$ in the $x$ and $y$ directions, and
$[-6H,6H]$ in $z$. Our standard resolution is $256\times256\times48$ (4
cells per $H$). We also experiment with a higher resolution of
$512\times512\times96$ (8 cells per $H$), 
and different size boxes (see Table \ref{tab:tab1}).

We found in our simulations that the code timestep was often limited
by the dynamics of low-density gas at the vertical
boundaries of our box, where pressure gradients were especially large.
To avoid catastrophic reductions in the 
timestep, we set a floor on the density of $10^{-4} \rho_0$.
Lowering this floor by an order of magnitude reduced the
timestep by at least 30\% but did not significantly
alter our results.

\subsection{Diagnostic Averages\label{sec:diag}}
To facilitate analysis and obtain statistical properties of our 3D
time-dependent flows,
we define a few ways to average physical variables.
The first is a volume (box) average:
\beq
\langle X \rangle \equiv \frac{\int \,X \,dx dy dz}{\int dx dy dz} \,. 
\label{eq:volavg}
\enq
We can also weight by density:
\beq
\langle X\rangle_{\rho} \equiv \frac{\int \, \rho X \, dx dy dz}{\int \, \rho \, dx dy dz} \, .
\label{eq:davg}
\enq
To see the height dependence of variables, we define a horizontal average using 
\beq
\langle X (z) \rangle \equiv \frac{\int \, X \, dx dy}{\int dx dy} \,.
\enq
Time averaging is denoted
\beq 
\langle X \rangle_t \equiv \frac{\int \, X \, dt}{ \int dt} \,.
\enq
Sometimes we will combine averages: e.g., $\langle \langle X \rangle \rangle_t$.

\section{RESULTS \label{sec:result}}
Results from our constant cooling time and optically-thin thermal
cooling experiments are given in \S\ref{sec:result_simple} and
\S\ref{sec:result_real}, respectively. Some time-and-space-averaged
properties of our simulations are listed in Table~\ref{tab:tab1}.

\subsection{Constant Cooling Time (Constant $\beta \equiv \Omega \tcool$) \label{sec:result_simple}}
We begin by presenting our standard $\beta=10$ run in \S\ref{sec:std}.
How the stress scales with cooling rate and height is discussed in
\S\ref{sec:alpha_tc}, and how turbulent (read: non-circular)
velocities depend on disk altitude is described in
\S\ref{sec:alpha_z}.

\subsubsection{Standard Run ($\beta = 10$) \label{sec:std}}

\begin{figure*}[!ht]
\includegraphics[width=0.4\textwidth]{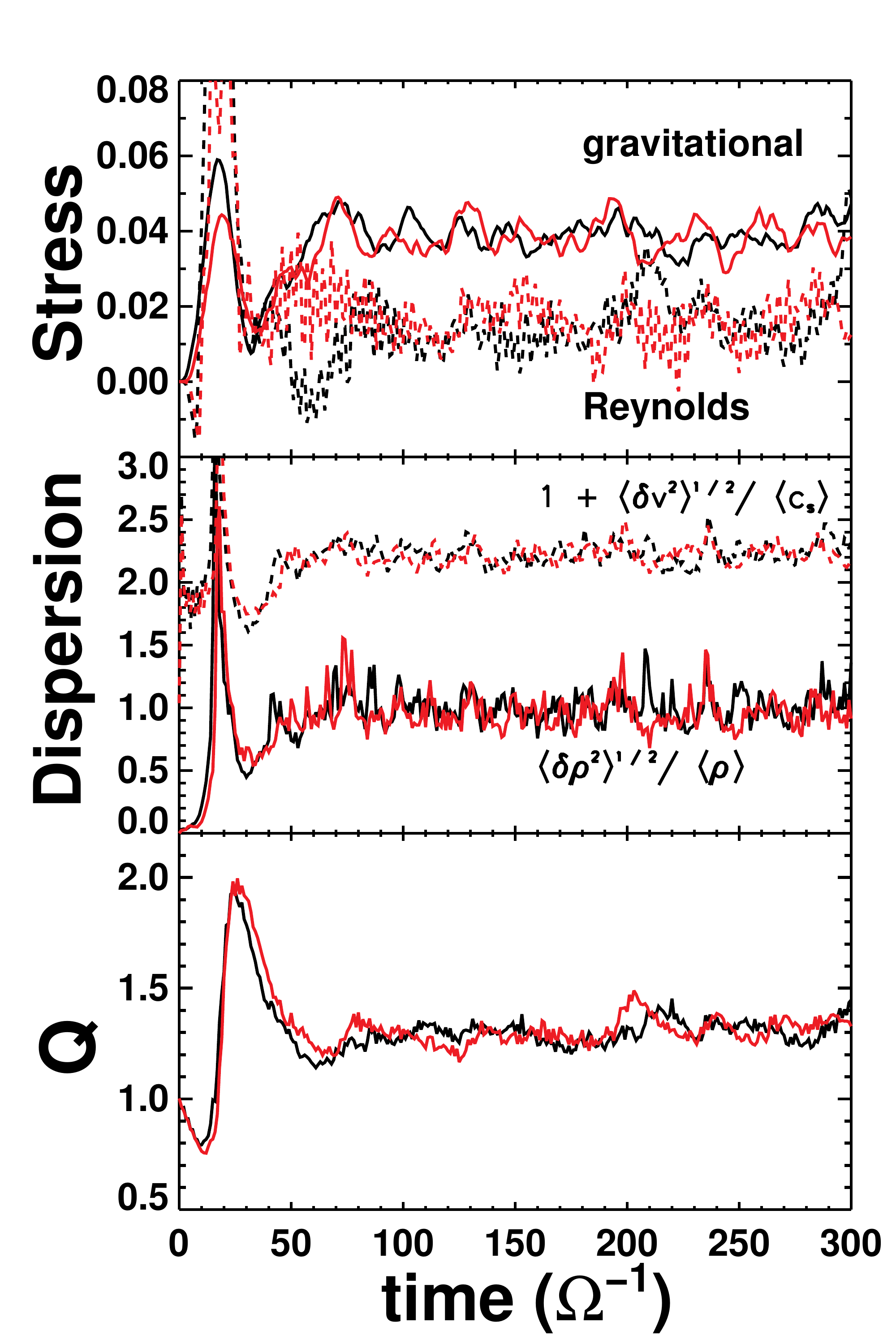}
\raisebox{.15\height}{\includegraphics[width=0.6\textwidth]{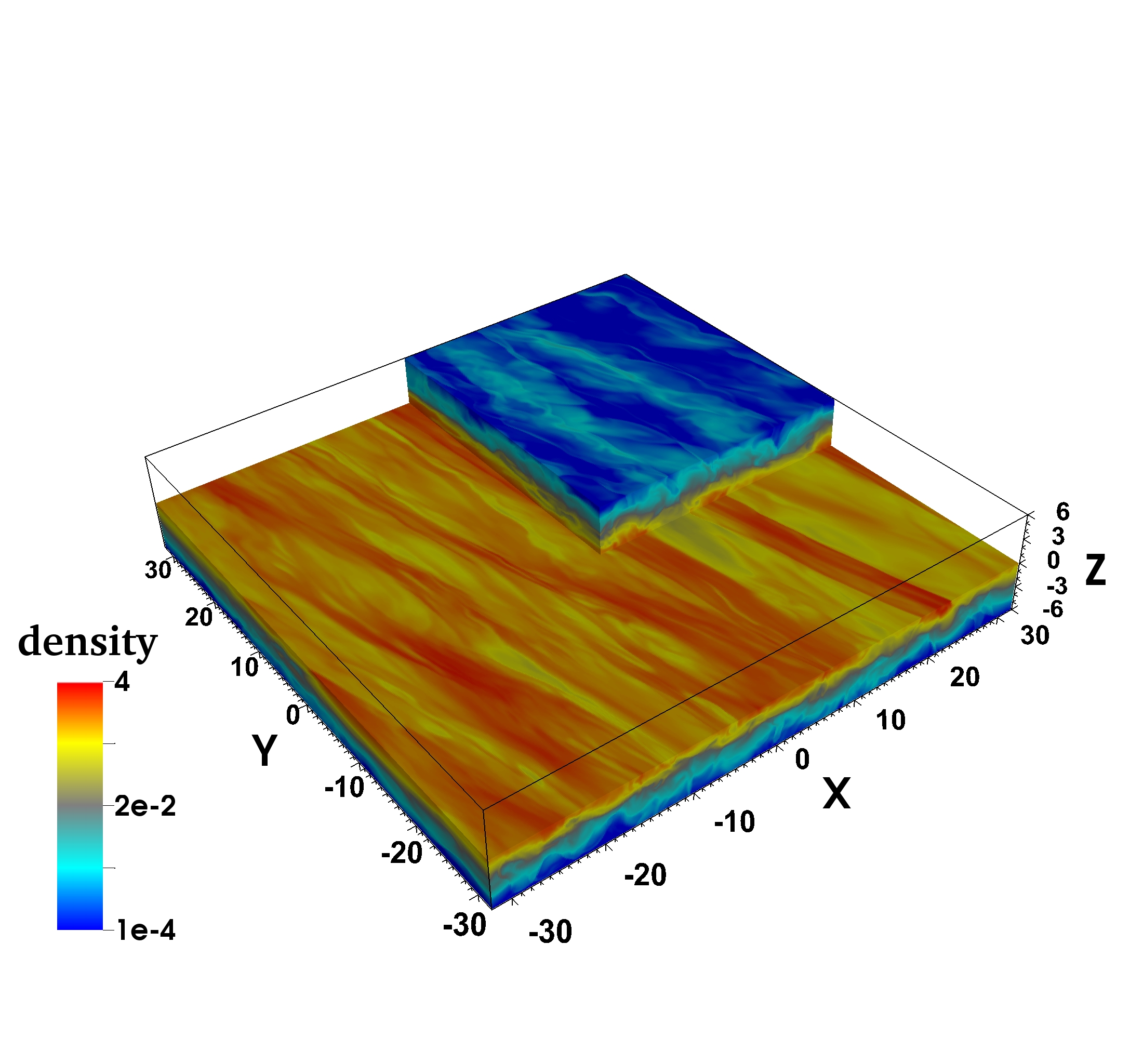}}
\caption{\small{Simulation results for constant $\tcool =
    10\Omega^{-1}$.  \textsl{Left}: Time histories of the Reynolds and
    gravitational stresses, normalized as in equation~(\ref{eq:shaksun}) and smoothed by
    $12\Omega^{-1}$ for clarity; volume-averaged rms
    density and velocity fluctuations (the latter is offset by 1 for
    clarity); and Toomre's $Q$. See text for definitions.  Black
    curves are drawn from our standard resolution simulation (tc=10)
    and red curves are from our high resolution simulation (tc=10.hi).
    \textsl{Right}: Cutaway view of density on a logarithmic scale at 
    $t=250\Omega^{-1}$ from
    our high-resolution simulation.}}\label{fig:std}
\end{figure*}

The simulations with $\tcool = 10 \Omega^{-1}$ (dubbed tc=10 at standard
resolution and tc=10.hi at high resolution in Table \ref{tab:tab1})
start from the hydrostatic equilibrium solution described in
section~\ref{sec:initial} and evolve for $30 \tcool = 300
\Omega^{-1}$.  The evolution of the disk as a whole can be followed by
constructing an effective Toomre's $Q$:
\beq
Q \equiv \frac{\langle c_{\rm s} \rangle_\rho \Omega}{\pi
  G\langle\Sigma\rangle} 
\enq 
where $\langle \Sigma \rangle$ is the vertical surface density
averaged horizontally and $\langle c_{\rm s} \rangle_\rho \equiv
\langle \sqrt{\gamma P / \rho} \rangle_\rho$ is the density-weighted,
box-averaged sound speed.

{\renewcommand{\arraystretch}{1.2}
\begin{deluxetable}{lccccccccc}
\tabletypesize{\footnotesize}
\tablecolumns{10} \tablewidth{0pc}
\tablecaption{3D Simulations of Gravito-Turbulent Disks\label{tab:tab1}}
\setlength{\tabcolsep}{0.05in}
\tablehead{
\colhead{\begin{tabular}{l} Name\tablenotemark{(a)}                    \\    \end{tabular} } &
\colhead{\begin{tabular}{c} $\Delta T$\tablenotemark{(b)}              \\ ($\Omega^{-1}$)   \end{tabular} } &
\colhead{\begin{tabular}{c} $\Delta T_{\rm avg}$\tablenotemark{(c)}    \\ ($\Omega^{-1}$)   \end{tabular} } &
\colhead{\begin{tabular}{c} $\langle\langle \tcool \rangle\rangle_{t}$       \\ ($\Omega^{-1}$)  \end{tabular}} &
\colhead{\begin{tabular}{c} $\langle\alpha_{\rm {Reynolds}}\rangle_{t}$\tablenotemark{(d)}  \\ $(\times 10^{-2})$  \end{tabular}} &
\colhead{\begin{tabular}{c} $\langle\alpha_{\rm {gravity}}\rangle_{t}$\tablenotemark{(d)}   \\ $(\times 10^{-2})$  \end{tabular}} &
\colhead{\begin{tabular}{c} $\aone$ \tablenotemark{(d)}                                     \\ $(\times 10^{-2})$  \end{tabular}} &
\colhead{\begin{tabular}{c} $\langle\alpha^{\prime}\rangle_{t}$ \tablenotemark{(d)}         \\ $(\times 10^{-2})$  \end{tabular}} &
\colhead{\begin{tabular}{c} $\langle\langle\delta v^2\rangle^{1/2}_{\rho}\rangle_{t}$ \\ $(H\Omega)$  \end{tabular}} &
\colhead{\begin{tabular}{c} $\langle\langle c_{\rm s} \rangle_{\rho}\rangle_{t}$  \\
$(H\Omega)$  \end{tabular}} 
}
\startdata
tc=3.hi \tablenotemark{(e)}   & $20$   & $-$     & $ 3$   &  $-$    & $-$ & $-$     &  $-$  & $-$  & $-$\\ 
tc=4.hi       & $200$ & $100$ & $ 4$   & $5.69$  & $6.50$ &  $12.2$ & $10.0$ & $2.31$  & $1.96$\\ 
tc=5.hi       & $200$ & $100$ & $ 5$   & $4.19$  & $5.82$ &  $10.0$ & $8.19$ & $2.20$  & $2.01$\\ 
tc=8.hi       & $200$ & $100$ & $ 8$   & $2.06$  & $4.41$ &  $6.47$ & $5.10$ & $1.90$  & $2.07$\\ 
tc=10         & $300$ & $100$ & $10$   & $1.88$  & $3.87$ &  $5.76$ & $4.24$ & $1.79$  & $2.18$\\ 
tc=10.dfl     & $300$ & $100$ & $10$   & $1.35$  & $3.99$ &  $5.33$ & $4.14$ & $1.71$  & $2.05$\\ 
tc=10.lxy128  & $300$ & $100$ & $10$   & $1.20$  & $4.03$ &  $5.23$ & $3.89$ & $1.98$  & $2.12$\\
tc=10.hi      & $300$ & $100$ & $10$   & $1.70$  & $3.82$ &  $5.52$ & $4.06$ & $1.79$  & $2.17$\\ 
tc=10.hi.dfl  & $300$ & $100$ & $10$   & $1.72$  & $3.95$ &  $5.67$ & $4.19$ & $1.74$  & $2.04$\\
tc=10.hi.lz10 & $100$ & $50$  & $10$   & $1.98$  & $3.78$ &  $5.77$ & $4.09$ & $1.73$  & $2.02$\\ 
tc=10.hi.lz14 & $100$ & $50$  & $10$   & $1.22$  & $4.03$ &  $5.25$ & $4.09$ & $1.74$  & $2.05$\\ 
tc=20         & $150$ & $100$ & $20$   & $0.29$  & $2.55$ &  $2.85$ & $2.14$ & $1.35$  & $2.17$\\ 
tc=20.hi      & $140$ & $100$ & $20$   & $0.46$  & $2.42$ &  $2.88$ & $2.02$ & $1.26$  & $2.07$\\  
tc=20.hi.dfl  & $140$ & $100$ & $20$   & $0.73$  & $2.47$ &  $3.20$ & $2.18$ & $1.31$  & $2.11$\\  
tc=40         & $300$ & $200$ & $40$   & $-0.06$ & $1.51$ &  $1.44$ & $1.06$ & $0.97$  & $2.10$\\ 
tc=40.hi      & $400$ & $200$ & $40$   & $0.27$  & $1.36$ &  $1.62$ & $1.03$ & $1.05$  & $2.14$\\  
tc=80         & $400$ & $200$ & $80$   & $0.13$  & $0.76$ &  $0.89$ & $0.55$ & $0.85$  & $2.26$\\ 
tc=80.hi      & $500$ & $200$ & $80$   & $0.35$  & $0.58$ &  $0.93$ & $0.52$ & $1.16$  & $2.26$\\  
tc=120.hi     & $500$ & $200$ & $80$   & $0.14$  & $0.39$ &  $0.53$ & $0.31$ & $1.09$  & $2.12$\\  
\hline 
b=100.hi \tablenotemark{(e)}   & $10$   & $-$     & $<3$   &  $-$    & $-$ & $-$     &  $-$  & $-$  & $-$\\ 
b=500.hi      & $140$ & $100$ & $9.64$ & $1.67$  & $3.99$ &  $5.66$ & $4.02$ & $1.61$  & $2.09$\\ 
b=500.hi.dfl  & $140$ & $100$ & $10.6$ & $1.53$  & $3.64$ &  $5.17$ & $3.64$ & $1.55$  & $2.07$\\ 
b=800.hi      & $240$ & $150$ & $17.1$ & $0.92$  & $2.56$ &  $3.47$ & $2.28$ & $1.31$  & $2.15$\\ 
b=1000.hi     & $240$ & $150$ & $19.4$ & $0.84$  & $2.29$ &  $3.13$ & $2.02$ & $1.28$  & $2.21$\\ 
b=2000.hi     & $240$ & $150$ & $28.2$ & $1.02$  & $1.66$ &  $2.68$ & $1.50$ & $1.24$  & $2.38$\\ 
b=3000.hi     & $240$ & $150$ & $45.8$ & $0.65$  & $1.00$ &  $1.66$ & $0.88$ & $1.05$  & $2.40$\\ 
\enddata 
\tablenotetext{(a)}{~Our naming convention is as follows: `tc=n' denotes a constant cooling time simulation
with $\beta=\Omega\tcool=$ n; `b=n' denotes a simulation with optically-thin thermal cooling with $b=$ n (see
\S\ref{sec:equation}); `.hi' implies a higher resolution of $512\times512\times96$ grid cells
instead of our standard $256\times256\times48$; `.dfl' simulations
use a density floor equal to $10^{-5}$ $\times$ our
initial midplane density (densities that fall below the floor
value are set equal to the floor value), whereas our standard
simulations use a floor value of $10^{-4}$ $\times$ the initial midplane
density; 
`.lxy128' denotes a wider (but not taller) box that spans $[-64H,64H]$ in the x and y directions instead of our standard $[-32H,32H]$;
and `.lz10' and '.lz14' denote box heights
of $10H$ and $14H$, respectively, instead of our
default height of $12H$.}
\tablenotetext{(b)}{~Duration of the simulation.}
\tablenotetext{(c)}{~Time span used for averaging various quantities, measured backward from the end of the run; e.g., a simulation of duration $\Delta T = 200 \Omega^{-1}$ for which $\Delta T_{\rm avg} = 100\Omega^{-1}$ takes its averaging interval between times $t_1 = 100 \Omega^{-1}$ and $t_2 = 200 \Omega^{-1}$.}
\tablenotetext{(d)}{~As defined by equations (\ref{eq:shaksun}) and (\ref{eq:alpha_prime}), respectively, $\alpha$
is the density-weighted average over the simulation domain, and $\alpha^{\prime}$ is the conventional
box averaged value with no density weighting. The total stress $\alpha = \alpha_{\rm Reynolds}+\alpha_{\rm gravity}$.}
\tablenotetext{(e)}{~For these runs, cooling is so rapid that the disk fragments gravitationally.}
\end{deluxetable}
}

Figure \ref{fig:std} displays the time evolution of $Q$.  The disk
settles into steady gravito-turbulence after an initial transient
phase lasting $\sim$50$\Omega^{-1}$.  The transient phase is violent:
the cooling disk collapses under its own weight during the first
$10\Omega^{-1}$; rebounds vertically as gas becomes strongly heated by
shocks from $t = 10\Omega^{-1}$ to $25\Omega^{-1}$; and relaxes from
$t = 25\Omega^{-1}$ to $50\Omega^{-1}$ into a quasi-steady state that
lasts the remainder of the simulation.  During the rebound phase,
about 10\% of the total mass in the box is lost through the vertical
boundaries.  This initial mass loss should be of no consequence as we are
interested in the final dynamical equilibrium attained by the disk,
i.e., the steady self-regulated state in which heating driven
by gravitational instability 
closely balances the imposed cooling.

The final equilibrium density profile is broader than
our initial profile, as illustrated in Figure \ref{fig:initial}.  In
the final state, mass
continues to be lost through the top and bottom of the box,
but at a rate so slow ($\lesssim$ 10\% from $t =
50$--$300\Omega^{-1}$) that 
we discern no obvious long-term trend in
any other statistical property that we measure.
This is demonstrated in Figure
\ref{fig:std}, which attests that in self-regulated gravito-turbulence
(for $\beta = 10$), $Q$
hovers around $1.33$; 
density fluctuations \beq \delta \rho = \rho
(x,y,z) - \langle \rho (z) \rangle \enq are on the order of unity {when normalized to the
average local density;
velocity fluctuations \beq \delta v = \sqrt{v_x^2 + \delta v_y^2 +
  v_z^2} \enq are mildly sonic (here $\delta v_y = v_y + 3\Omega x/2$
is the non-Keplerian azimuthal velocity); and the Shakura-Sunyaev
stress-to-pressure parameter
 \beq \label{eq:shaksun} \alpha \equiv \frac{{\langle w_{xy}
    \rangle}_\rho}{\langle P \rangle_\rho} \equiv \frac{\langle
  \newton + \reynolds \rangle_\rho}{\langle P \rangle_\rho} \enq
fluctuates about 0.055, with the gravitational stress ($\newton$) 
exceeding the Reynolds stress ($\reynolds$) by roughly a factor of 3 
(here $g_x \mathbf{\hat{x}} + g_y \mathbf{\hat{y}} + g_z \mathbf{\hat{z}}$ is the local
gravitational acceleration).
Our definition for $\alpha$
weights by density; it tries to ``follow the mass'' in our
stratified simulations and should be more accurate, e.g., when
applied to 2D studies that evolve $\Sigma$ instead of $\rho$. In the
literature---which reports on simulations that are commonly
unstratified---the stress-to-pressure ratio is more often calculated 
as a volume average without weighting by density
\citep[e.g., equation~19 in][]{gammie01}:
\beq
\label{eq:alpha_prime} 
\alpha^{\prime} \equiv \left|\frac{d\ln{\Omega}}{d\ln{R}}\right|^{-1}\!\! \frac{\int\! w_{xy} dxdydz}{\int\!\rho
c_{\rm s}^2 dxdydz}
 = \frac{2}{3\gamma}\, \frac{\langle w_{xy}\rangle}{\langle P \rangle} 
\enq
where $R$ is disk radius. 
This conventional $\alpha^{\prime}$ is $0.5$--$0.8$ times
our $\alpha$ in constant cooling time simulations.
We calculate both measures of stress in this paper; see, e.g.,
Table~\ref{tab:tab1}.
Our results appear robust insofar
as doubling the resolution in every direction changes $\alpha$ by just
a few percent (compare red and black curves in Figure \ref{fig:std},
and see also Table \ref{tab:tab1}).

\subsubsection{Stress vs.~Cooling Rate and Height: $\alpha (\tcool, z)$
 \label{sec:alpha_tc}}

\begin{figure}[!h]
  \epsscale{1.0}
\plotone{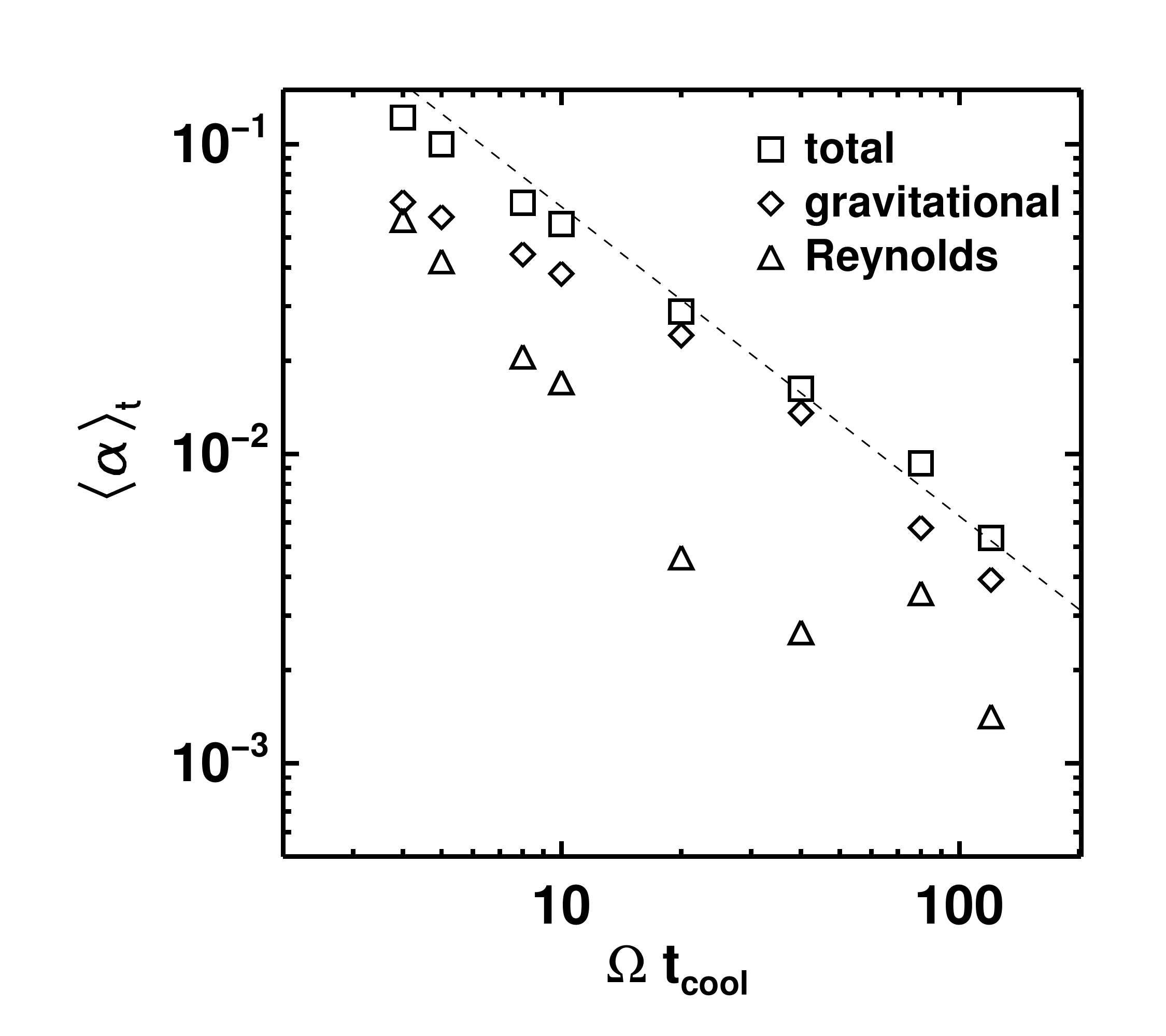} \caption{\small{Density-weighted 
time-and-space-averaged stresses $\alpha$ (as defined in equation~\ref{eq:shaksun}) vs.~cooling time for our high-resolution
constant $\beta$ simulations. 
The dashed line scales as $1/(\Omega\tcool)$
and is shown for reference. See also Figure \ref{fig:a_tc_gammie}
which plots the volume-averaged $\alpha^\prime$ without weighting by density.
}}
\label{fig:a_tc}
\end{figure}

All simulations were run for long enough that we can recover
    steady curves at late times like those shown in
    Figure~\ref{fig:std}, enabling us to evaluate meaningful time
    averages.  In only one constant cooling run, tc=3.hi ($\beta =
    3$), did the disk fragment instead of settling into steady
    gravito-turbulence (see also our analogous optically-thin run
    b=100.hi for which $\beta < 3$, which also fragmented).  Thus we
    establish (for $\gamma=5/3$) that the collapse criterion in 3D
    local disks is $\tcool \lesssim 3\Omega^{-1}$.  This is consistent
    with the fragmentation boundary reported previously in 2D local
    \citep[e.g.,][]{gammie01} and 3D global studies
    (e.g., \citealt{CHL2007}; but see 
    \citealt{MB2011}, \citealt{MB2012}, and \citealt{Riceetal2014}
    for cautionary remarks.
Figures~\ref{fig:a_tc} and \ref{fig:a_tc_gammie} show
how the density-weighted $\langle \alpha \rangle_t$ and the volume-averaged
$\langle \alpha^\prime \rangle_t$---and their 
gravitational and hydrodynamic components---vary with $\tcool$ for our
high-resolution runs (time averages are taken over intervals
lasting at least $100 \Omega^{-1}$; see Table \ref{tab:tab1}). 
Gravitational stresses are observed to exceed hydrodynamic streses by factors of
$\sim$2--5.
We find that $\alpha$ and especially $\alpha^\prime$ 
scale closely with $1/(\Omega\tcool)$, as
expected from energy conservation \citep{gammie01}.
Figure~\ref{fig:a_tc_gammie} for $\alpha^\prime$
reveals that our simulation results 
match the analytic expectation
\beq
\alpha^{\prime} = \frac{4}{9\gamma(\gamma-1)} \frac{1}{\Omega\tcool}
\label{eq:a_tc}
\enq
nearly perfectly for our chosen $\gamma=5/3$, indicating excellent
energy conservation in our code.

\begin{figure}[!h]
  \epsscale{1.0}
  \plotone{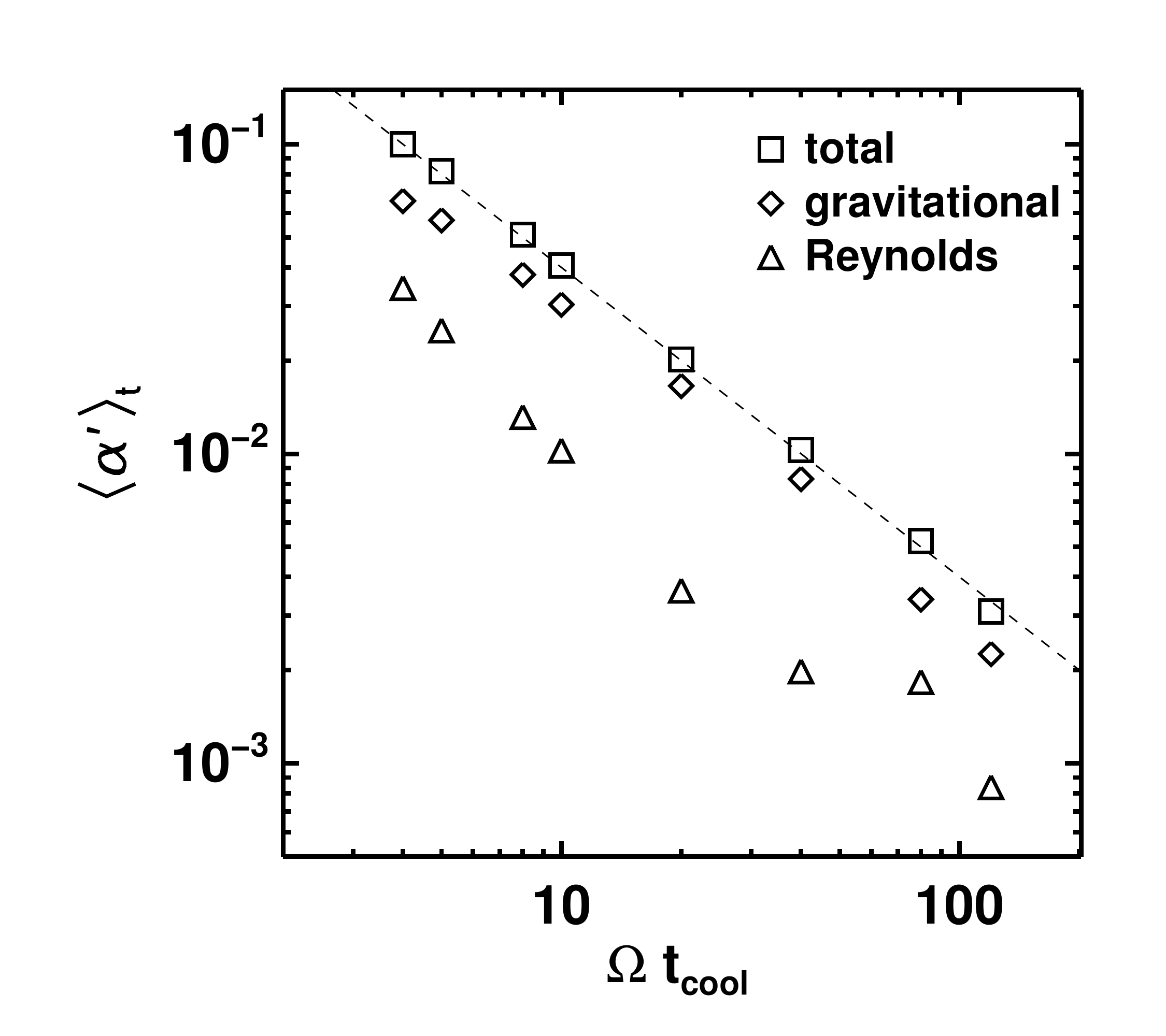} 
  \caption{
   \small{The conventional time-and-volume-averaged
   stresses $\alpha^\prime$ (see equation~\ref{eq:alpha_prime}, which is equivalent to
   equation~19 of \citealt{gammie01})
   vs.~cooling time for our high-resolution
   constant $\beta$ simulations. 
   The dashed line shows the analytic prediction of
   equation~(\ref{eq:a_tc}) (see also equation~20 of 
   \citealt{gammie01} for the case of a 2D disk), evaluated for our
chosen $\gamma = 5/3$. The simulation results
   match the analytic formula, implying that our code
   conserves total energy well.}
   }
\label{fig:a_tc_gammie}
\end{figure}

\begin{figure}[!h]
\epsscale{1.0}
\plotone{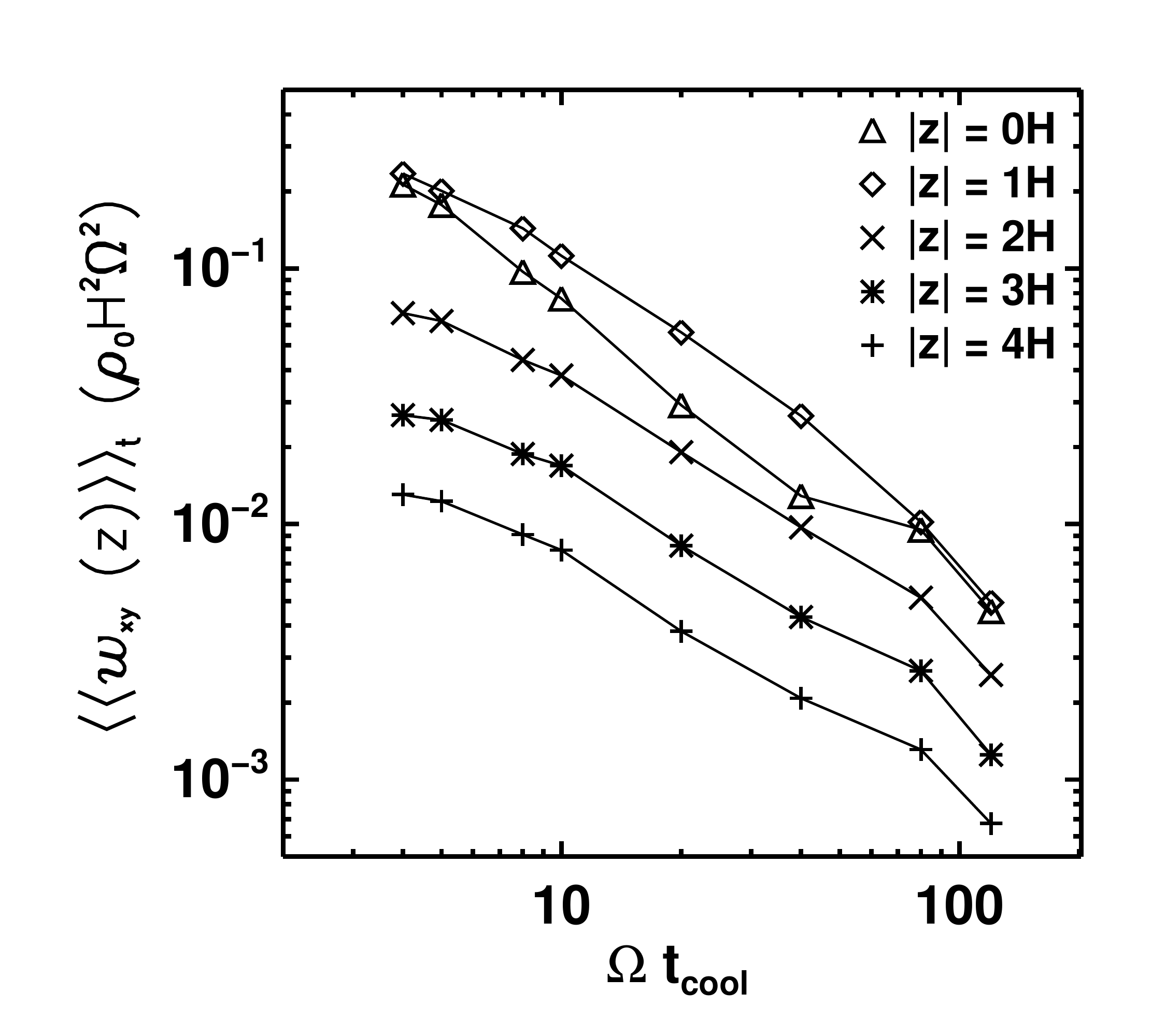}
\caption{\small{Time-averaged total absolute stress (not normalized by
    pressure) as a function of cooling time for our high-resolution
    constant $\beta$ runs, for various disk heights. In making this
    plot, we average data above and below the midplane, and over a
    time interval lasting $>2\tcool$ at the end of each simulation
    (see Table \ref{tab:tab1}). Stresses tend to decrease with height,
    but only by factors of $\sim$6--16 from $z=0$ to $|z|=4H$, while the
    density drops by nearly $10^3$.  
    In Figure \ref{fig:cum_stress} we
    will see that the gravitational stress at altitude is exerted
    largely by density fluctuations at the midplane. The time-averaged
    stress actually peaks slightly away from the midplane because the
    Reynolds stress at the midplane alternates in sign periodically,
    leading to a partial cancellation in the time average there. 
    }}
\label{fig:a_tc_z}
\end{figure}

A similar scaling of stress with $\tcool$ 
applies at all heights, as demonstrated in Figure \ref{fig:a_tc_z}.
Note how the local stress---which is dominated by gravitational
forces---diminishes by factors of $\sim$6--16 from $|z|=H$ to $|z|=4H$, or roughly
as $|z|^{-n}$ with $n=1.2$--$2$,
by contrast to the density $\rho$ which drops exponentially with height
(Figure \ref{fig:initial}). The comparatively slow drop-off 
of stress with height arises because gravity is a long-range
force: the stress at altitude arises from gravitational
forces exerted by density fluctuations near the midplane. We illustrate
this in Figure \ref{fig:cum_stress} by plotting the contribution
to the total gravitational stress exerted at $z=4H$ 
from all heights. About $80\%$ of the stress at $z=4H$ comes from the
gravity of the disk between $-H < z < H$.

\begin{figure}[!h]
  \epsscale{1.0} \plotone{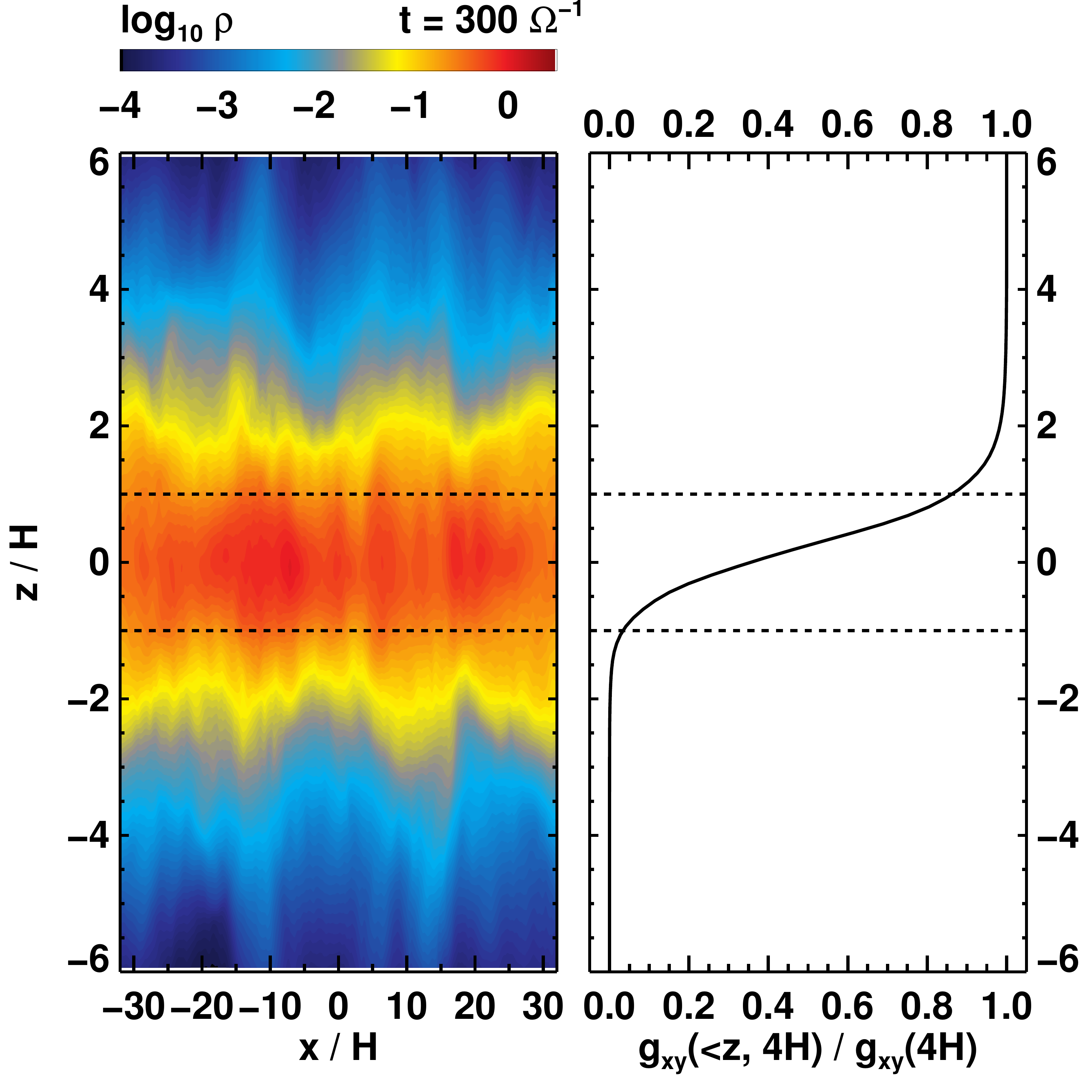}\caption{\small{
      Contributions from all heights to the gravitational stress $g_{xy} \equiv
      g_xg_y/4\pi G$ at $z=+4H$. 
      \textsl{Left}: Poloidal snapshot of the azimuthally
          averaged log density at $t=300\Omega^{-1}$ from our $\beta=10$
          high resolution run.
          \textsl{Right}: the running fraction of the gravitational
          stress at $+4H$ from heights less than $z$; by definition,
      this fraction is zero for $z = -6H$ (there is no disk material
      below the bottom of our box) and unity for $z = 6H$ (all
      material is contained below the top of our box). Dashed
          lines show where $z=\pm H$.  The lion's share of the
      stress at $+4H$ originates from between $z=-H$ and $z=+H$, where
      more than $75\%$ of the disk mass resides.}}
\label{fig:cum_stress}
\end{figure}

Total $\alpha$-values generally change by 10\% or less when the
resolution is doubled in all directions; when the imposed density
floor is lowered from $10^{-4}\rho_0$ to $10^{-5}\rho_0$; when the
box size is doubled horizontally; or when the box height varies from
$12H$ to $10H$ or $14H$ (Table \ref{tab:tab1}).
Increasing the
spatial resolution tends to increase the Reynolds stress and lower the
gravitational stress.  
Velocity fluctuations are better resolved with
finer grids, but the gains are modest; doubling the resolution in all 
directions changes non-circular velocities by $\sim$30\% at most and 
typically by only several percent;
at the same time, the increased Reynolds stress should be
offset by a decrease in gravitational stress, since the total heating
rate must balance the imposed cooling rate (see Figure \ref{fig:a_tc_gammie}). 
Note that simulations with larger $\beta$ (longer cooling times)
have smaller amplitude fluctuations and are more computationally
demanding---this may be the reason why, e.g., our $\beta=80$
runs exhibit more sensitivity to resolution than our $\beta < 80$ runs (see Table \ref{tab:tab1}).
Unless indicated
otherwise, the data plotted in all our figures and discussed in the
text are drawn from our high resolution simulations with a standard
density floor of $10^{-4}\rho_0$ (i.e., the .hi models in Table
\ref{tab:tab1}).

\subsubsection{Turbulent Velocities vs.~Height\label{sec:alpha_z}}

\begin{figure}[!h]
  \epsscale{1.1} \plotone{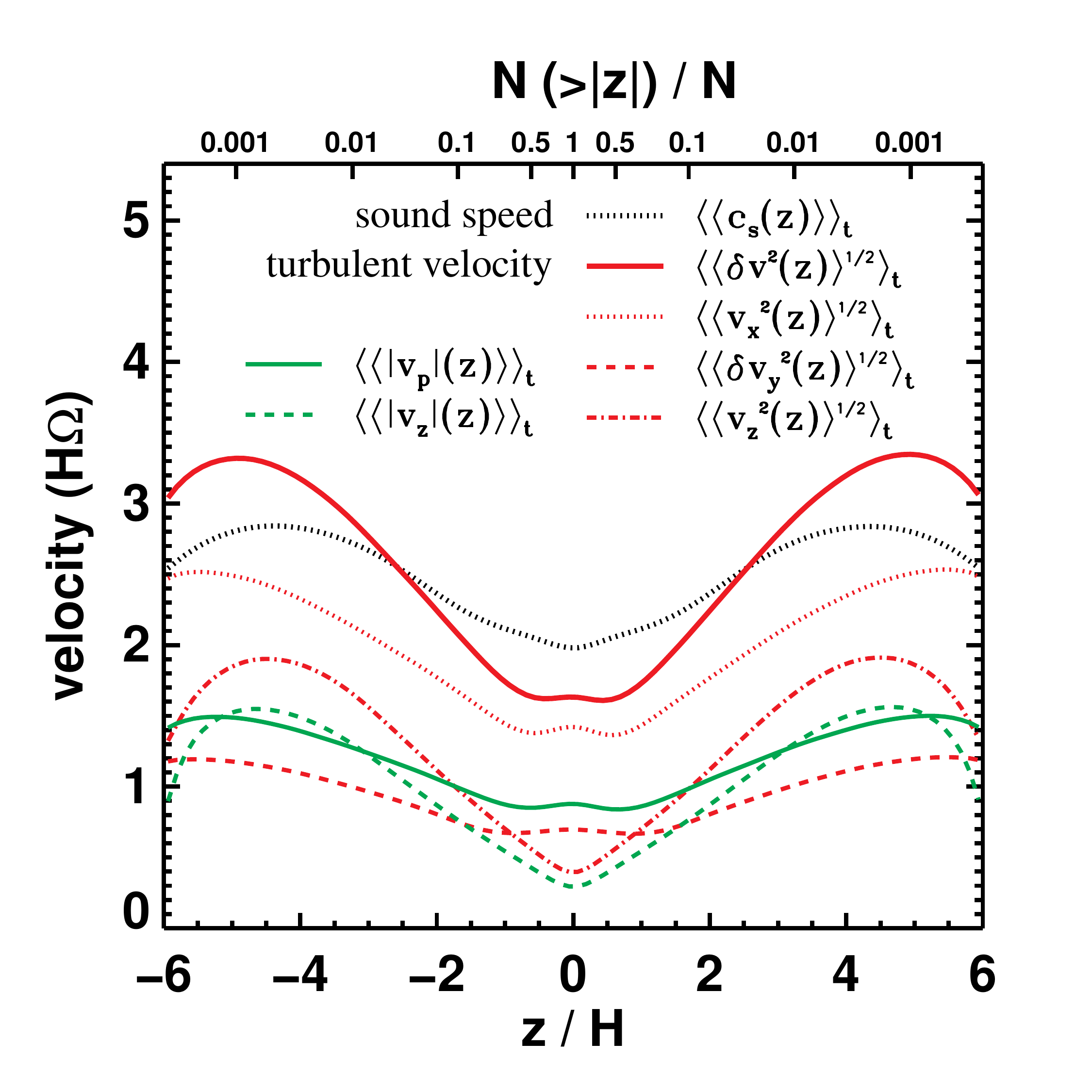} \caption{\small{Vertical
      profiles of the sound speed and turbulent (read: non-circular)
      velocities, time-averaged from $t=200$--$300\Omega^{-1}$ using
      our high-resolution $\Omega \tcool = 10$ run (tc=10.hi). Neither
      the sound speed nor the turbulent velocities vary much with
      height, even as the overlying column density above a given
      height (measured on the top x-axis) drops by three orders of
      magnitude.  Red curves denote individual components of rms
      turbulent velocities, while green curves denote our proxies for
      line-of-sight turbulent velocities, as defined in
      \S\ref{sec:alpha_z}. Here and elsewhere, we discount the behavior
      at $|z| \gtrsim 5H$ as it is sensitive to our vertical box boundary
      conditions. By contrast, results at $-5H < z < 5H$ are far enough
      away from the vertical boundaries to be robust; see Figure \ref{fig:3box}.}}
\label{fig:prof_vel}
\end{figure}

The variation with height of the rms value of the turbulent velocity
$\delta v$
(i.e., the non-Keplerian bulk motion) is shown in Figure
\ref{fig:prof_vel} for $\beta=10$. Turbulent velocities vary only
weakly with altitude, increasing by less than a factor of two from
midplane to surface as the overlying column density drops by more than
three orders of magnitude. 
That turbulent motions are just as
vigorous at altitude as they are at depth might at first glance seem
surprising, since densities at altitude fall below the Toomre
density $\Omega^2/2\pi G \sim 0.5$ \citep{sc2013}. In other words, gas at
altitude is not self-gravitating and does not generate turbulence in
and of itself. But in hindsight, the relatively flat velocity profile
is understandable, because as we noted in section \ref{sec:alpha_tc},
gravity is long-range: density fluctuations at the midplane---where gas
is of Toomre density and is self-gravitating---exert gravitational
forces at altitude and strongly perturb the rarefied gas there.

In fact, non-Keplerian motions are, if anything, greater at altitude
than at depth.  
Vertical velocities $v_z$ vary more strongly
with height (factor of 5 increase from midplane to
surface) than do in-plane velocities $v_{\rm p}$,
  \footnote{As shown in Figure \ref{fig:prof_vel},
  the drop in the vertical component of turbulent velocity
  from $|z|\approx 5H$ to $6H$ in our standard box is a
  numerical artifact, as we have discovered by varying the box
  height to $10H$ (run tc=10.hi.lz10) and $14H$ (tc=10.hi.lz14)---see
  Figure~\ref{fig:3box}.  For any given box, the total rms turbulent velocity
  rolls over at a distance of $\sim$1$H$ 
  from either vertical boundary. The roll-over might be due to the enforced 
  outflow boundary conditions that artificially eliminate any inward 
  motion. Conversely, our results away from the vertical boundaries
  appear to have converged with box size.\label{foot:artifact}}
which is sensible because the vertical gravitational acceleration from
disk self-gravity tends to increase away from the midplane: at the
midplane, vertical forces from the upper and lower disk tend to
cancel.
\begin{figure}[!h]
\epsscale{1.1}
\plotone{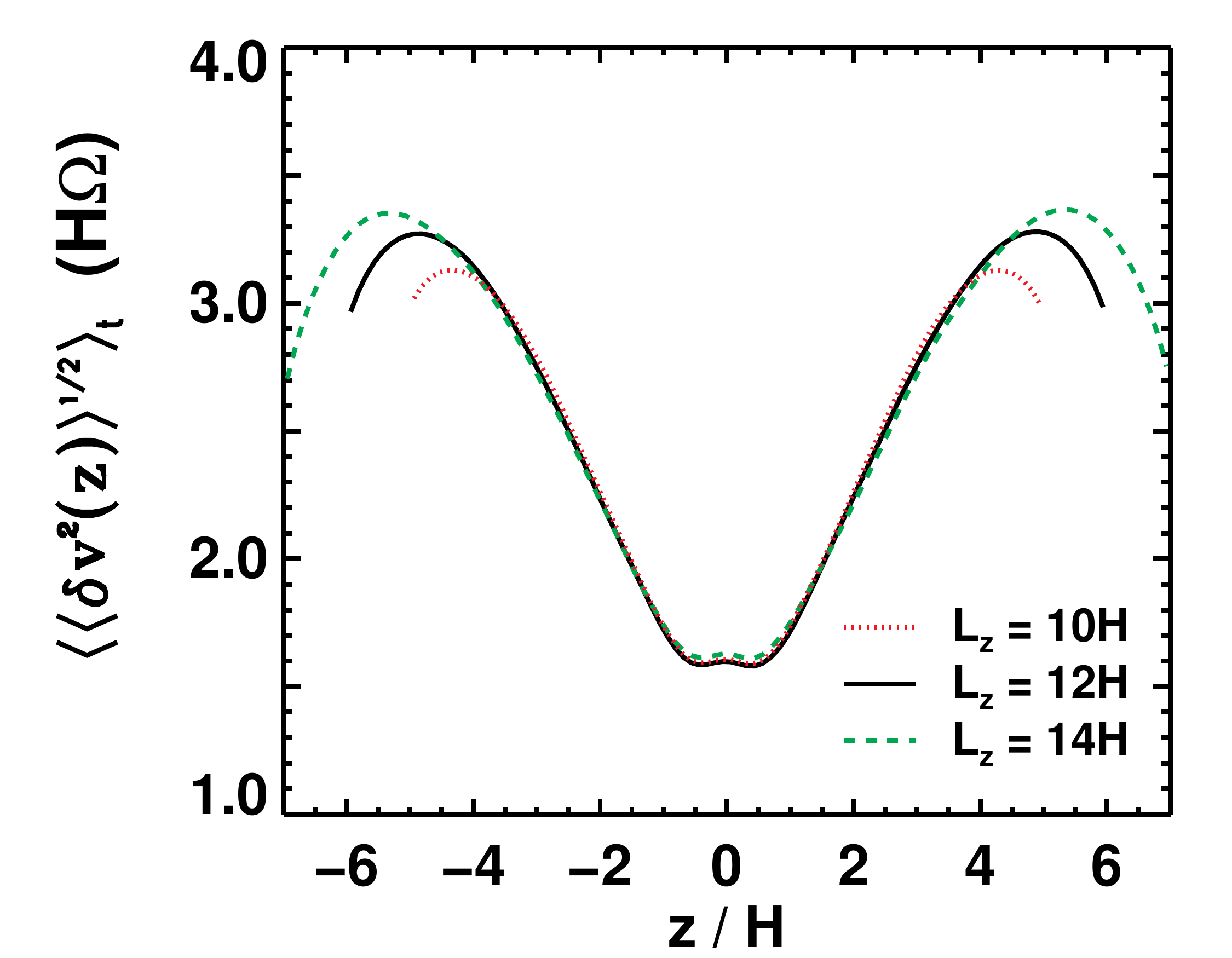}
\caption{\small{Vertical profiles of rms turbulent velocity 
	        $\langle\langle \delta v^2(z)\rangle^{1/2}\rangle_t$ for
		simulations with different box heights $L_z$. 
		The fact that in all simulations, velocities roll
		over $\sim$$1H$ away from box boundaries
		indicates that the velocity behavior there
                is spurious and subject to artificial boundary effects.
                Consequently, throughout this paper, when we discuss our results
                for our standard box having $L_z = 12H$, we
                restrict our statements to $|z| \lesssim 5H$.}}
\label{fig:3box}
\end{figure}
In addition, fluid motions in all directions should amplify at
altitude from the steepening of waves launched upward from the
midplane and which propagate down density
gradients. \citet{Simon2011} suggested from their numerical
experiments that increasing velocity dispersions with height are
generic to turbulent disks, and our results are consistent with this
proposal. At $|z| > H$, motions are strongest in the $x$-$z$ plane,
driven by self-gravity in those directions; azimuthal perturbations
are suppressed by Keplerian differential rotation which shears
overdensities into nearly axisymmetric structures. We see evidence for
a kind of $x$-$z$ circulation, whereby gas compresses radially from
self-gravity and spurts out vertically, as illustrated in Figure
\ref{fig:xzmotion}.

\begin{figure}[!h]
\epsscale{1.1}
\plotone{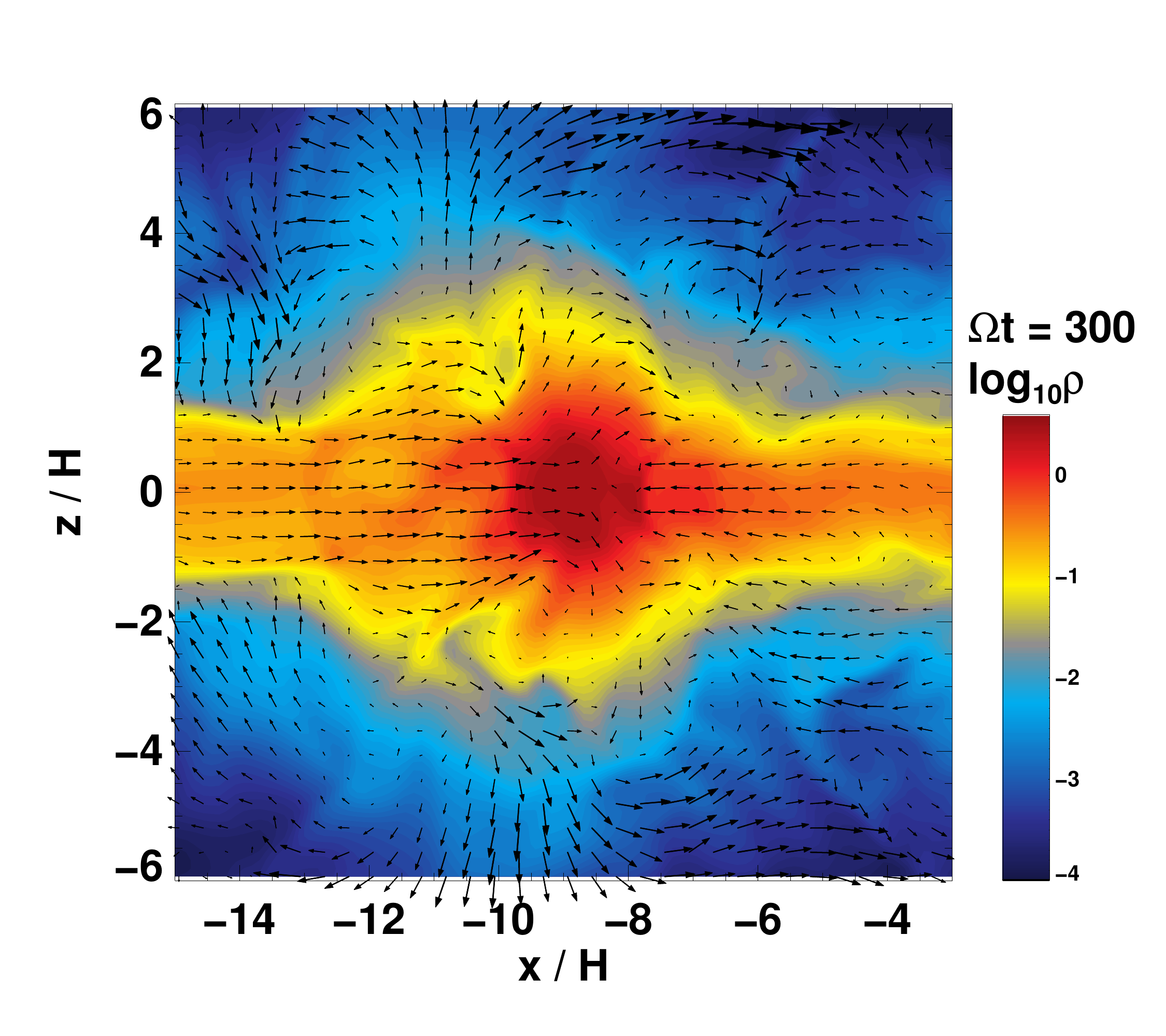}
\caption{\small{A meridional snapshot, taken at $t = 300\Omega^{-1}$
    and fixed azimuth $y=0$, of our high-resolution $\Omega \tcool=10$ run
    (tc=10.hi).  Colors show log density and arrows show the velocity
    field. The longest arrow has an $x$-$z$
    velocity of $8.17 H\Omega$, or $1.45$
    of the local sound speed. These velocities are within a few percent
    of the total 3D non-circular velocities, as contributions
    from $\delta v_y$ are typically small.}}
\label{fig:xzmotion}
\end{figure}

We also calculate velocity dispersions more closely related to
observables. We forego a full spectral line analysis and instead compute
a line-of-sight (los) turbulent velocity in the disk plane, averaged
over azimuth $\phi$:
\beq
|v_{\rm p}| \equiv \frac{1}{2\pi}\int^{2\pi}_{0} d\phi \left|v_{\rm x} \cos\phi - \delta v_{\rm y} \sin\phi
\right| 
\label{eq:vlos}
\enq
(\citealt{Simon2011}; \citealt{Forgan2012}).
Here we use $\delta v_{\rm y}$ for the turbulent velocity in the
azimuthal direction, not to be confused with the bulk velocity $v_{\rm y}$ which 
includes the background shear velocity.
The quantity $|v_{\rm p}|$ is a proxy for the
actual los turbulent linewidth for disks seen edge-on. For disks
seen face-on, the corresponding proxy velocity is $|v_z|$.  Figure
\ref{fig:prof_vel} plots the vertical profiles for $\langle |v_{\rm
  p}(z)| \rangle$ and $\langle |v_z (z)|\rangle$ (each horizontally
averaged), while the top left panel of Figure \ref{fig:pdf} 
displays the full probability
density functions for $|v_{\rm p}|/c_{\rm s}$ and $|v_z|/c_{\rm s}$
for several cuts in height, all for $\beta=10$. There is little
variation in these proxy los velocities with height, particularly for
$|v_{\rm p}|$. The most probable value of $|v_{\rm p}|/c_{\rm s}$ 
increases by a factor of only $1.2$ from
$z > 0$ to $z > 4H$ (for $\beta=10$), while
$|v_{z}|/c_{\rm s}$ increases by a factor of $2.5$.
We obtain similarly flat velocity profiles for all $\beta > 3$ runs; the only
difference is a systematic shift
toward smaller velocities as $\tcool$ increases, e.g., for $\beta = 10
(40)$, the most probable value of $|v_{\rm p}|/c_{\rm s}$ is $0.35
(0.22)$, sampled over all $z > 0$; see
Figure~\ref{fig:pdf}).\footnote{\citet{Forgan2012} also reported
  in their global simulations of self-gravitating disks that turbulent
  velocity distributions are virtually unchanged from $z>0$ to $z>H$;
  see their Figure 3, and note that their turbulent velocities and
  stresses are considerably lower than ours, presumably because the
  cooling times of their disks are much longer. They observed that the
  linewidth probability distributions sampled at different heights were
  ``remarkably similar, but the poor resolution of higher altitudes
  forbids us from attributing this to any phenomenology of the disc.''
  By comparison, our local simulations are well resolved vertically
  and show that the flatness of the turbulent velocity profile extends
  more than three orders of magnitude in column density from midplane
  to disk surface.}

Note that the variations in Mach numbers between simulations with
different cooling times are almost all due to variations in 
turbulent
velocities, not to variations in the sound speed. Table \ref{tab:tab1}
indicates that the sound speeds of various simulations are all about
the same---as expected, since the input surface densities and self-regulated
Toomre $Q$-values are all about the same, irrespective of cooling time.
Prolonging the cooling time
reduces turbulent velocities but does not alter background
temperatures (at fixed surface density).

\begin{figure*}[!ht]
\includegraphics[width=0.5\textwidth]{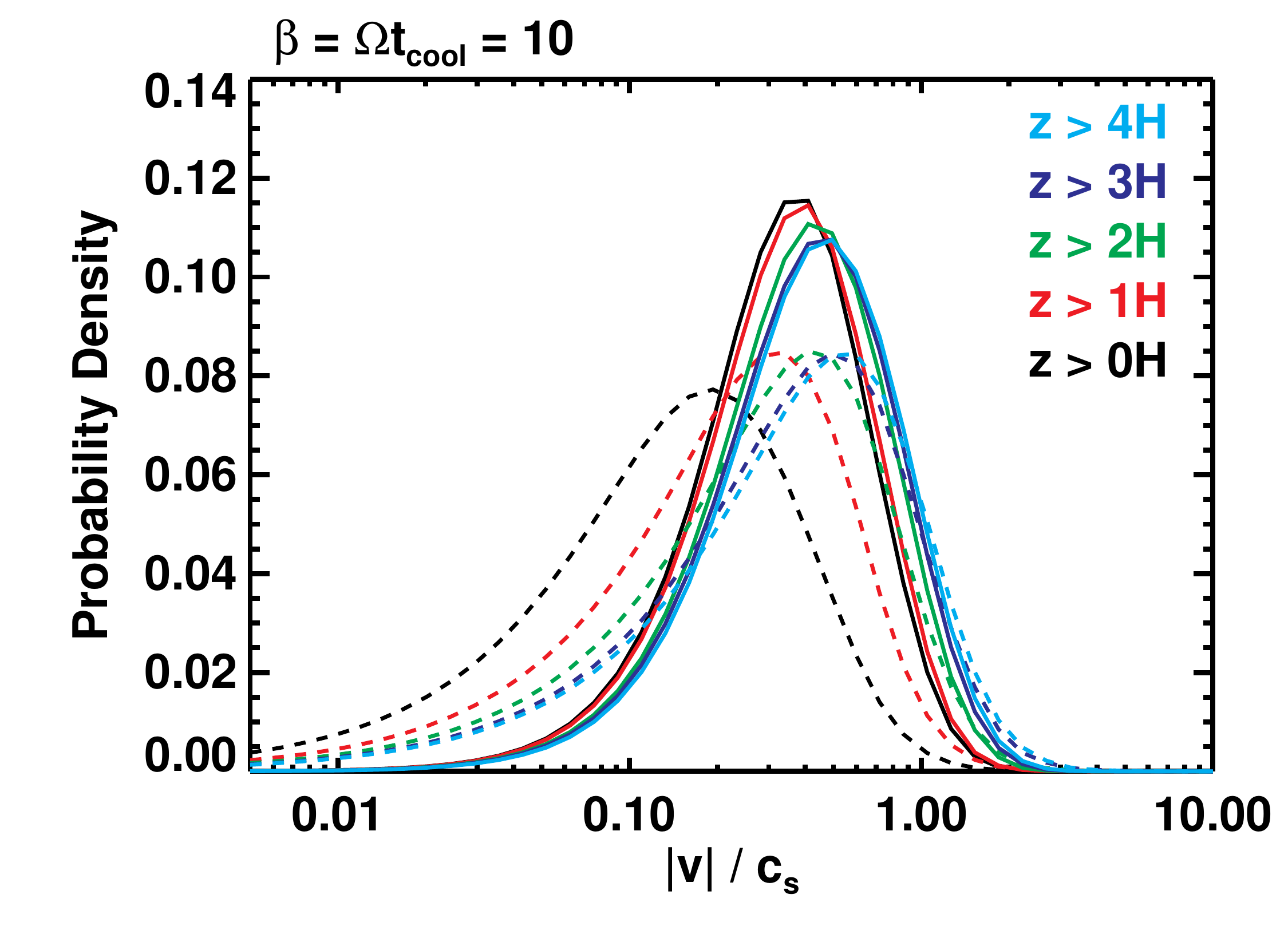}\hfill
\includegraphics[width=0.5\textwidth]{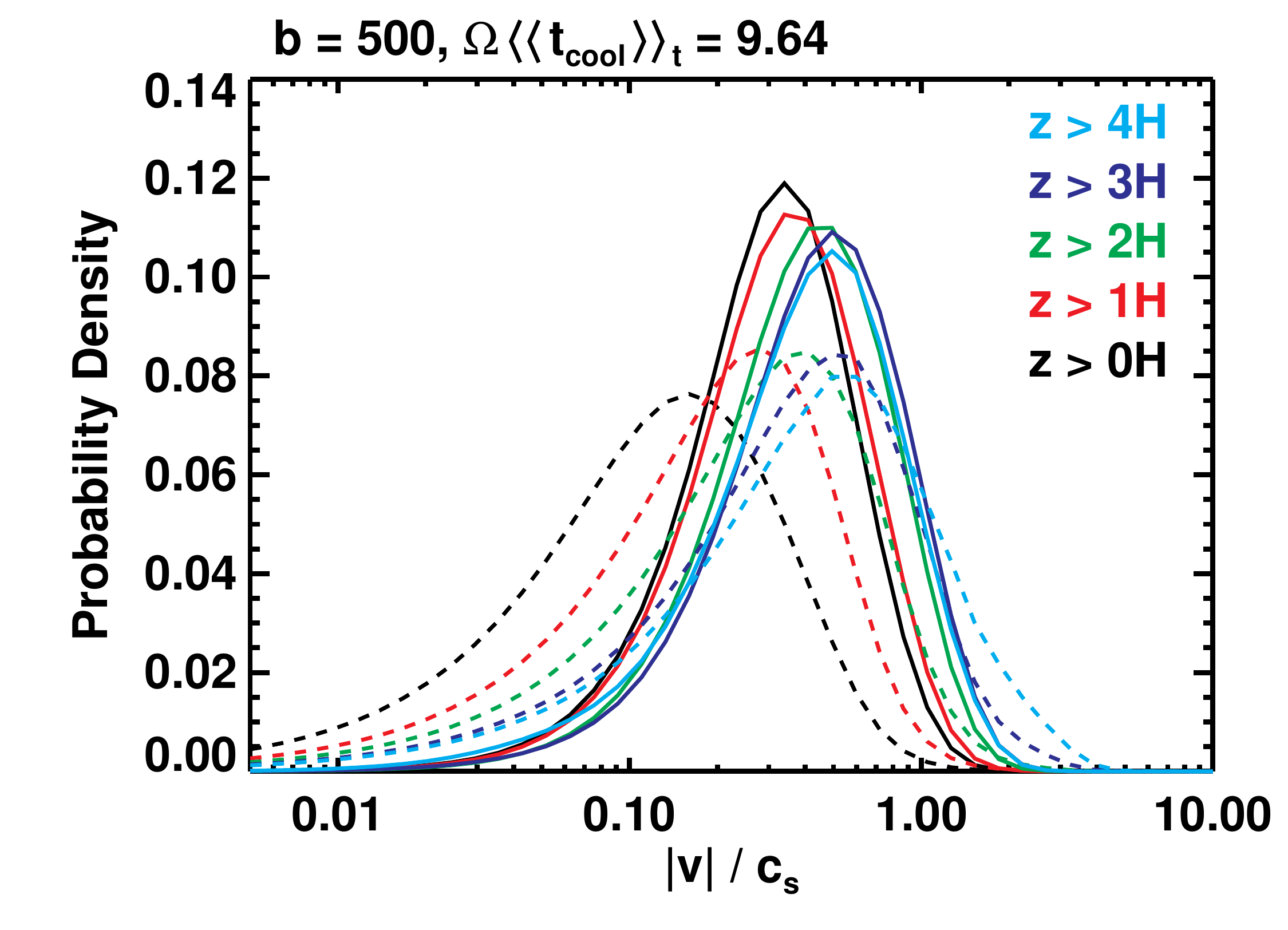}\hfill
\includegraphics[width=0.5\textwidth]{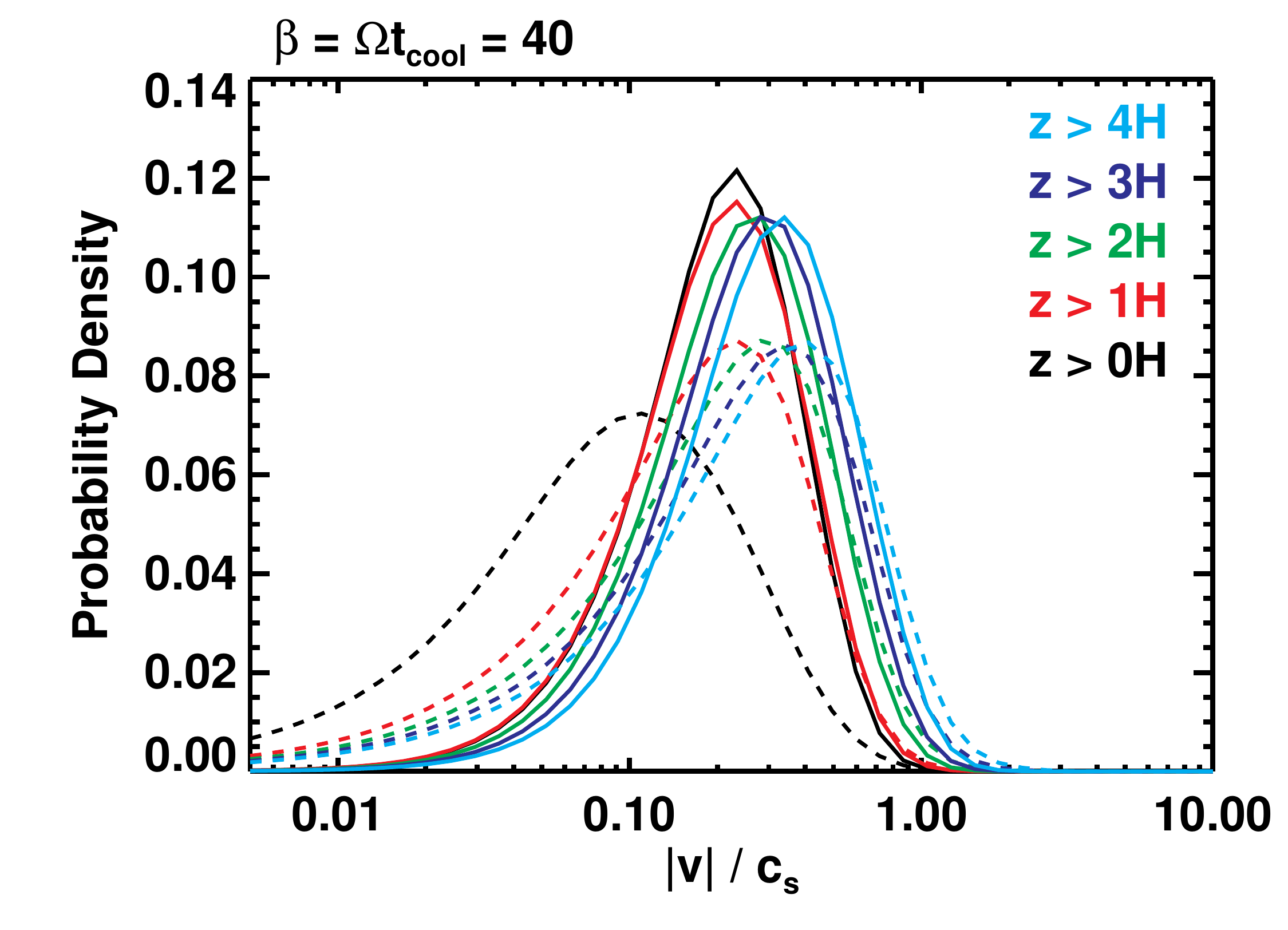}\hfill
\includegraphics[width=0.5\textwidth]{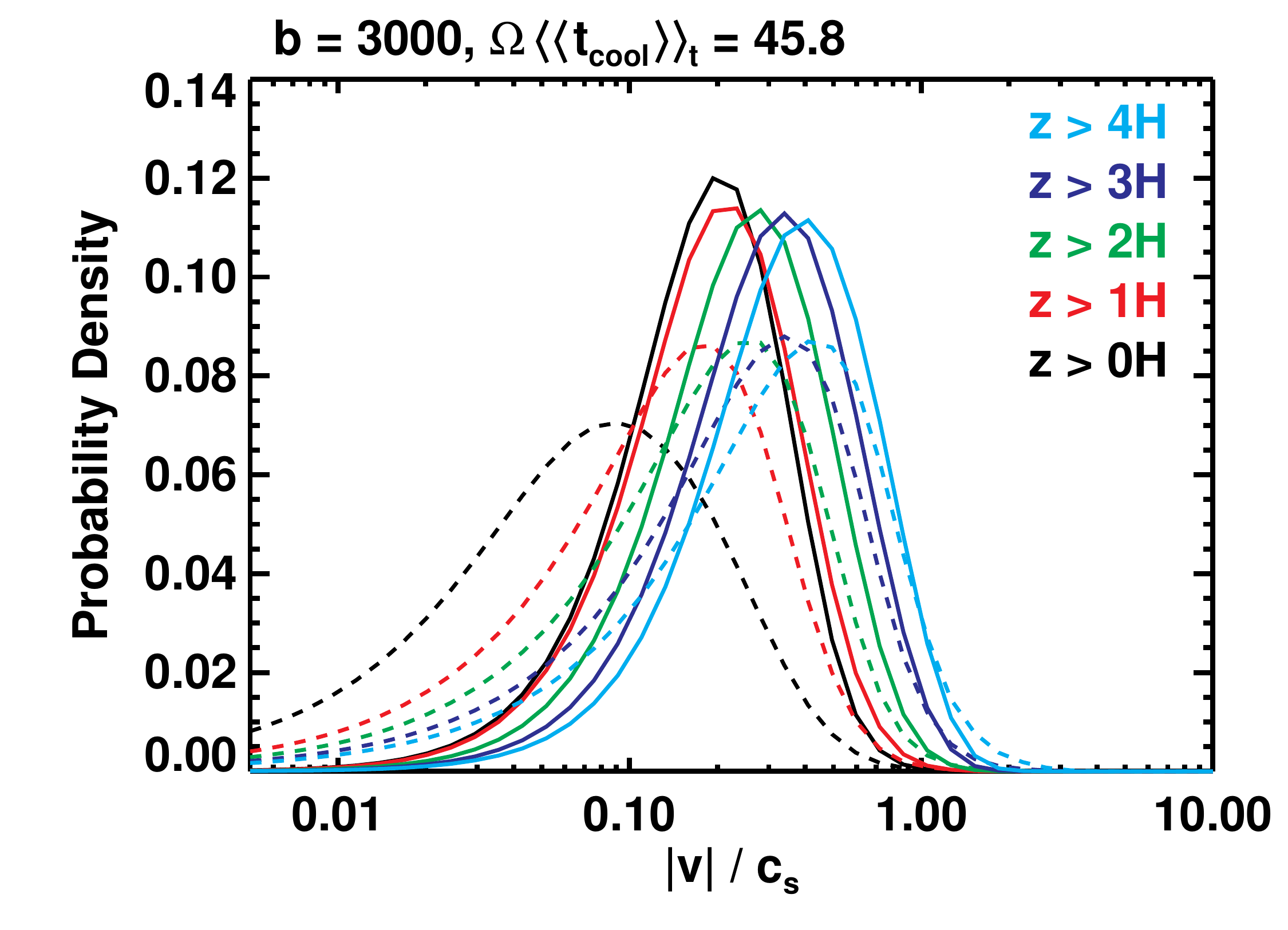}\hfill
\includegraphics[width=0.5\textwidth]{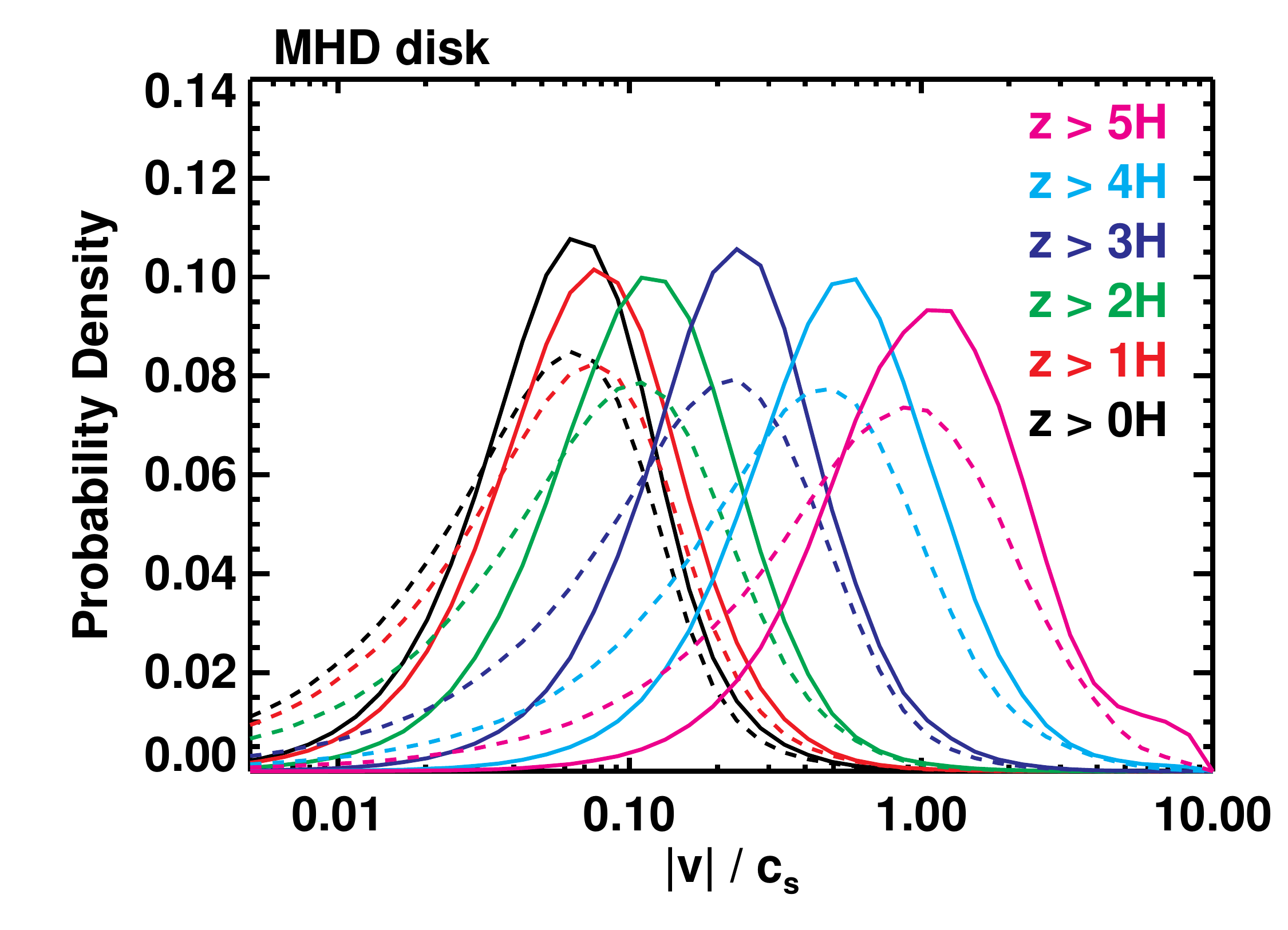}\hfill
\includegraphics[width=0.5\textwidth]{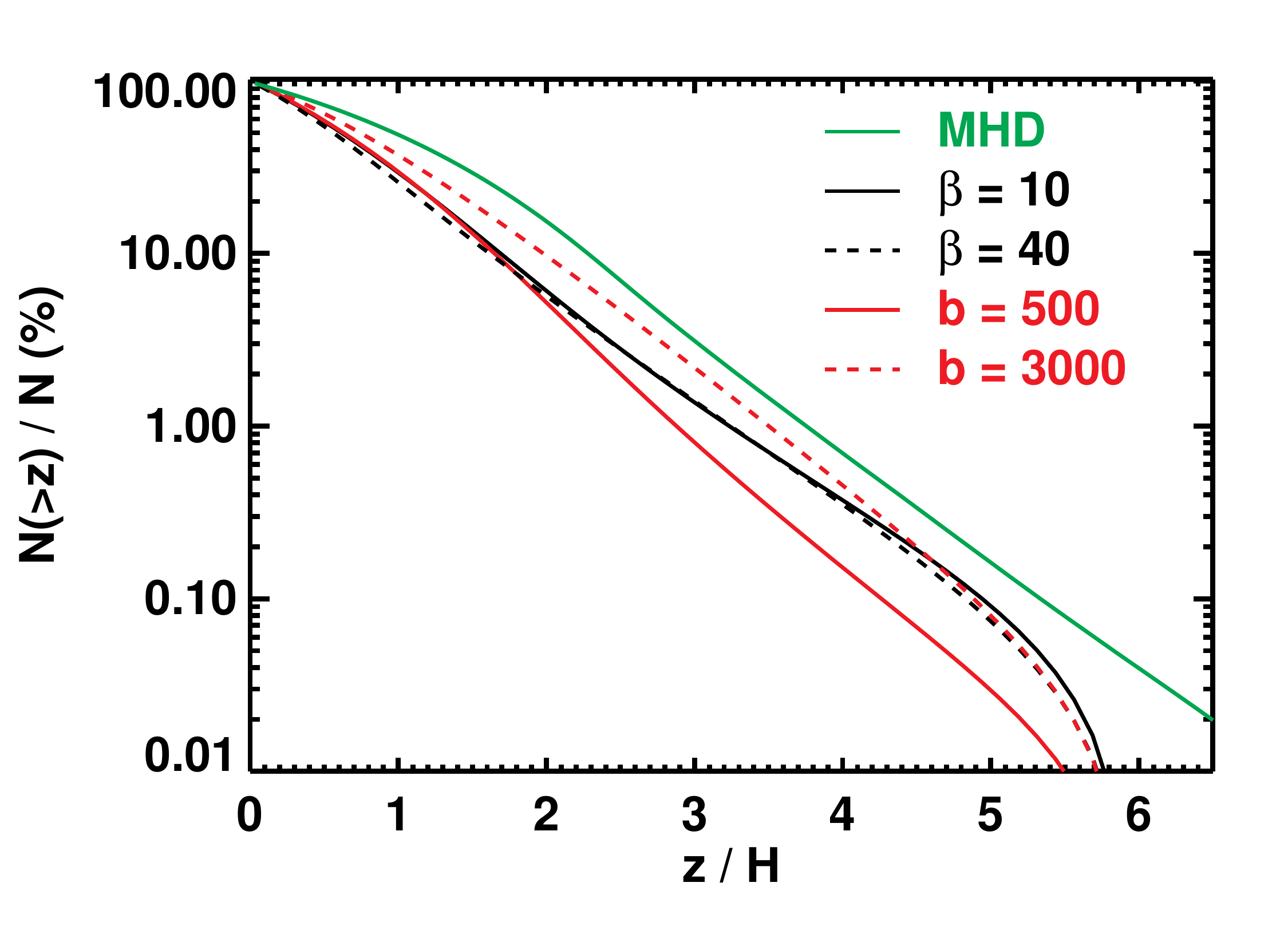}
\caption{\small{ 
    Probability density distributions, sampled at various heights, of in-plane 
    turbulent Mach numbers ($|v_{\rm p}|/c_{\rm s}$, solid curves) and vertical
    turbulent Mach numbers ($|v_{\rm z}|/c_{\rm s}$, dashed curves),
    for constant $\beta \in \{10,40\}$
    and optically-thin $b \in \{500,3000\}$ simulations, as labeled.
    Probability densities are probabilities per unit log Mach number.
    Also shown for comparison are results from an MRI-turbulent
    disk, run STD32 of \citet{Shi2010}. 
    How the fractional overlying column density $N(>z)/N$ varies with height $z$
    for each model is shown at bottom right.
    Gravito-turbulent velocities
    vary less with height than do MHD-turbulent velocities.
}}
\label{fig:pdf}
\end{figure*}

Figure \ref{fig:pdf} also makes a head-to-head comparison between the
velocity distributions of our gravito-turbulent disks and those of a
disk made turbulent by the magneto-rotational instability (MRI).  Our
MRI data are extracted from the standard run (STD32) of
\citet{Shi2010}, who studied the MRI using a stratified,
non-self-gravitating, ideal MHD shearing box with zero net magnetic
flux (for details, see their sections 2 and 3.1).  
Turbulent velocities vary more strongly with height in MRI-turbulent disks than
in gravito-turbulent disks; as $z$ increases from the midplane and the
overlying column density drops by three orders of magnitude, both
$|v_{\rm p}|/c_{\rm s}$ and $|v_z|/c_{\rm s}$ increase in the MRI-turbulent disk
by factors of $\sim$15, from $\sim$0.06 to $\sim$1.
This variation characterizes ideal MHD disks. 
Similar results, including supersonic velocities at $z > 3 H$,
were obtained by \citet{Simon2011}.
In non-ideal disks (those affected by Ohmic dissipation, ambipolar diffusion,
and/or the Hall effect), the variation
of magnetically driven velocities with altitude tends to be even steeper 
(\citealt{Simon2011,Simon13b}; see also Figure 7 of \citealt{lesur14}).

\citet{Forgan2012} suggested that the ratio $|v_{\rm p}|/|v_{\rm z}|$
could be used to differentiate between gravito-turbulent and
MRI-turbulent disks.  They found that $|v_{\rm p}| \sim 6 |v_{\rm z}|$
near the midplanes of their gravito-turbulent disks, whereas MHD
turbulence is more isotropic (even at the midplane
of a dead zone, $|v_{\rm p}|$ differs from  $|v_{\rm z}|$ by
a factor $\lesssim 3$ according
to the top panel of Figure 4 of \citealt{Simon2011}). 
Our results are nominally of
higher resolution, and show that $|v_{\rm p}|$ is indeed higher than
$|v_{\rm z}|$, but only by factors of $2$--$3$ at the midplane (Figure
\ref{fig:pdf}). Unfortunately, even this difference appears to vanish
at altitude. Thus the usefulness of $|v_{\rm p}|/|v_{\rm z}|$ in
distinguishing between gravito-turbulence and MRI turbulence appears
limited.

\subsection{Optically-Thin Thermal Cooling \label{sec:result_real}}

\begin{figure}[!h]
\epsscale{1.0}
\plotone{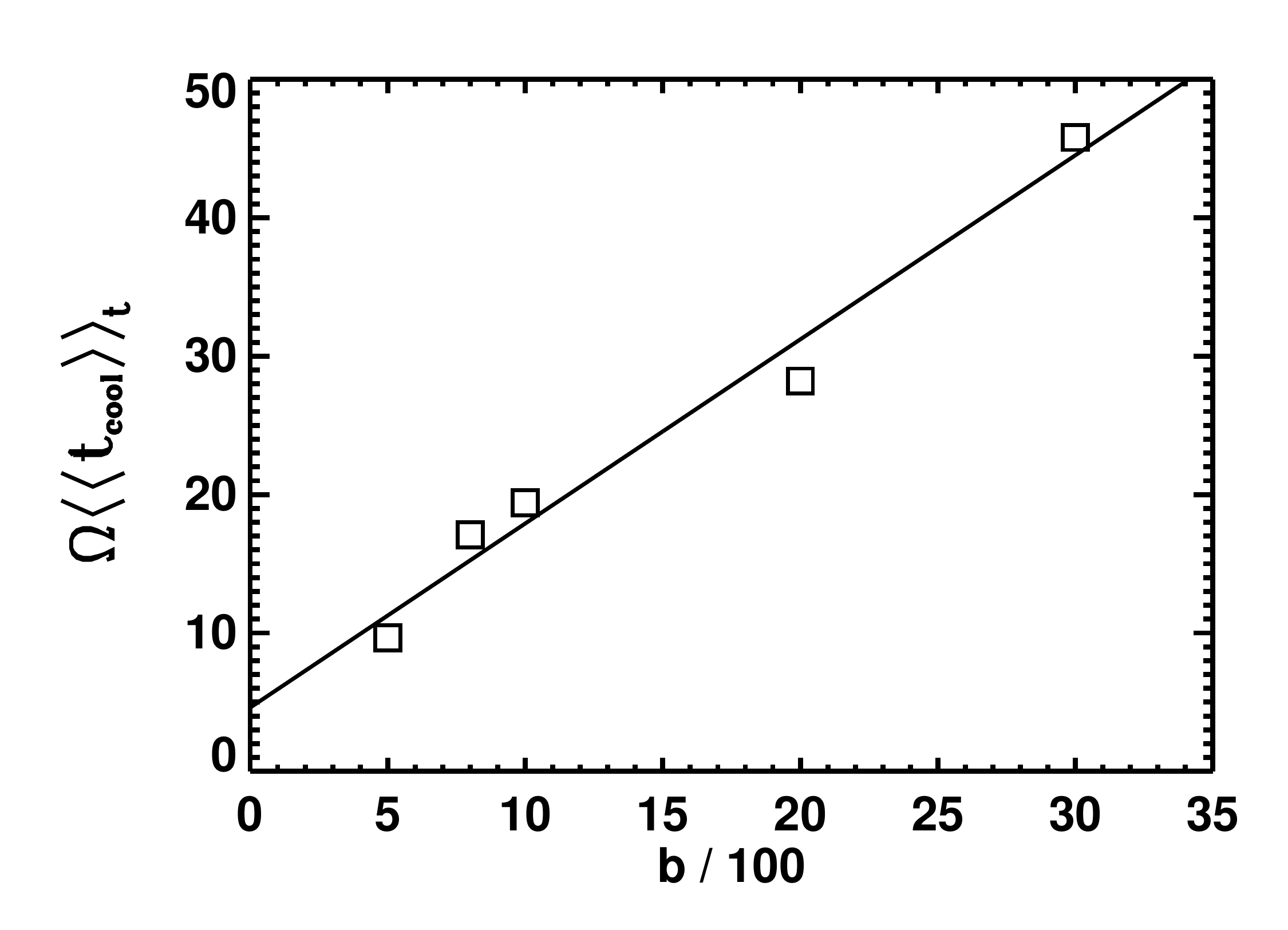}
\caption{\small{Relation between the $b$-parameter 
    (equation \ref{eq:qloss}) of our
    optically-thin thermal cooling experiments, and the effective
    cooling time (equation \ref{eq:tc_eff}).  Squares
    represent simulation results, while the solid line is the best
    linear fit ($\Omega \tceff = 0.0113 b + 4.61$, with $b$ in code
    units). The linear relation reflects the fact that
    our adopted cooling rate scales strongly with $T$ (as $T^4$)
    and tends to thermostat disks to the same temperature, regardless
    of $b$ and regardless of height.
}}
\label{fig:b_tc}
\end{figure}

Initial conditions for each of our optically-thin cooling experiments
(section \ref{sec:equation}) are taken from our constant cooling time
simulation tc=10.hi at $t=160 \Omega^{-1}$. Optically-thin cooling is
turned on immediately, with the cooling parameter $b \in \{100, 500,
800, 1000, 2000, 3000\}$ in code units; see equation (\ref{eq:qloss}).  
For every $b$, we can compute
an effective cooling time
\beq \tceff \equiv \frac{\langle\langle
  U\rangle\rangle_{t}}{\langle\langle \rho\qloss
  \rangle\rangle_{t}} \,.
\label{eq:tc_eff}
\enq
Figure \ref{fig:b_tc} shows that $b$ scales
linearly with $\tceff$.  Our optically thin disks quickly
settle into new gravito-turbulent states, except in the $b=100$ run,
where the disk cools so rapidly ($\tceff \lesssim 3\Omega^{-1}$) that it
fragments. For $b > 100$, we recover the nearly inverse-linear scaling
between stress and cooling time (see Table \ref{tab:tab1}).
When computing averages, we sample data after the disk
settles into steady gravito-turbulence, and use time intervals
lasting $\sim$100--150 $\Omega^{-1}$ or $\gtrsim 3$--$10\tceff$. 

In general, our optically-thin thermally cooling disks behave similarly to our
constant $\tcool$ disks. For example, 
$\langle\alpha\rangle_{t}= 0.0566$ for $b=500$ ($\tceff = 9.6
\Omega^{-1}$), and $\langle \alpha \rangle_t =0.0552$ for $\beta=10$
($\tcool = 10 \Omega^{-1}$). For both prescriptions, cooling is local;
furthermore, since our optically thin disks have cooling times that
depend only on temperature --- i.e., $\tcool \propto b/T^3$ --- their
cooling times are effectively constant as long as the disks are nearly
isothermal.  
In fact, Figure \ref{fig:prof_vel_b} shows that our
optically thin disks are more uniform in temperature than our constant
cooling time disks at $|z| < 4H$, presumably because the cooling flux
increases rapidly as $T^4$ and thermostats the gas.
\begin{figure}[!h]
\epsscale{1.1}
\plotone{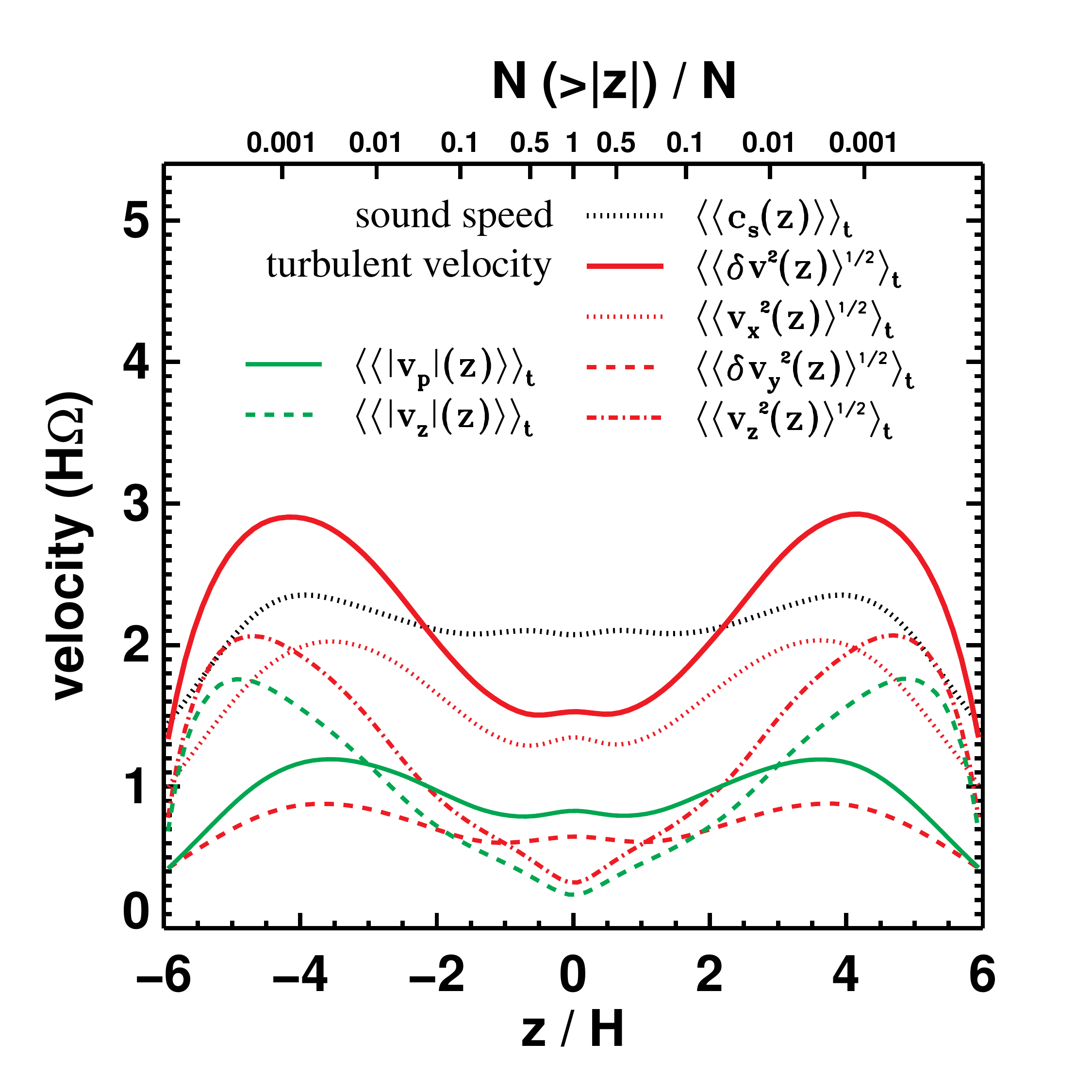}
\caption{\small{Turbulent velocity and sound speed profiles
for our simulation using optically-thin thermal cooling with $b=500$
(run b=500.hi).
As was the case in our constant cooling time experiments (Figure \ref{fig:prof_vel}), turbulent velocities and sound speeds change only modestly
with height.
Here and elsewhere, we ignore our results at $|z| > 5H$
because of numerical boundary effects; by contrast,
the behavior at $-5H < z < 5H$ is robust to changes in 
box size.
}}\label{fig:prof_vel_b}
\end{figure}

As was the case for our constant $\beta$ disks, turbulent velocities
of optically thin disks vary modestly with height. For example,
in-plane velocities increase by a factor of 1.5 from
$|z|=0$ to $4H$, and vertical velocities increase by a factor
of 5, even while the density drops by three orders of magnitude.
Probability distributions for the in-plane and vertical Mach numbers
are displayed in Figure \ref{fig:pdf}.  These distributions shift
toward smaller Mach numbers as $b$ increases (i.e., as the opacity
decreases and the cooling time increases): the most probable value of
$|v_{\rm p}|/c_{\rm s}$ (sampled over all $|z|>0$) changes from $0.31$ for $b
= 500$ to $0.2$ for $b=3000$.

\section{SUMMARY AND DISCUSSION \label{sec:conclusion}}
We used local shearing box simulations to study optically thin,
cooling, gravito-turbulent disks. Our investigation is basically the
3D version of Gammie's (\citeyear{gammie01}) study.  Our main result
is that non-circular motions driven by gravito-turbulence are nearly
independent of height above the midplane, across density contrasts as
large as $10^3$. The insensitivity of turbulent velocity to height is
due to gravity's long-range nature.  Although gas at altitude is too
rarefied to be itself self-gravitating, it is strongly gravitationally
accelerated by Toomre-density fluctuations near the midplane.  This
behavior contrasts with that in disks made turbulent by the
magneto-rotational instability (MRI).  In MRI-unstable disks,
non-circular velocities increase by more than a factor of 10 from
midplane to surface, and velocity variations are even larger when
non-ideal magnetohydrodynamic effects are present.
Our theoretical results, coupled with empirical
measurements of non-thermal line broadening in protoplanetary disks,
promise to distinguish between gravito-turbulence and MRI
turbulence---or perhaps to rule out both in favor of some other
transport mechanism (\citealt{Hughes2011}; Hughes et al.~2014, in
preparation).

Our results were derived under the assumption that the disk is
optically thin to its own cooling radiation. We obtained similar
simulation outcomes by assuming either a fixed cooling time or a
cooling time that depends on the local temperature.  To avoid
fragmentation, these cooling times must exceed the local dynamical
time.  Optically thin, self-gravitating disks with cooling times
longer than their dynamical times may not be too hard to find in
nature, at least among protoplanetary disks.\footnote{
Disks in active galactic nuclei become gravitationally
unstable at such large radii and where surface densities
are so low that their local cooling times are too short
to maintain quasi-steady gravito-turbulence (\citealt{goodman2003};
but see \citealt{BL2001} for an alternative view).}
At an orbital distance of $R \sim 100$ AU from a solar-mass star, a
surface density of $\Sigma \sim 10$ g/cm$^2$ (i.e., a disk mass of
$\Sigma R^2 \sim 0.01 M_\odot$) and a temperature of $T \sim 10$ K
leads to a Toomre $Q$ on the order of unity. The optical depth and
cooling time depend on the dust opacity $\kappa$, which is notoriously
uncertain, as it depends on the grain size distribution and the
dust-to-gas ratio.  At the sub-millimeter wavelengths characterizing
most of the cooling radiation from the outer disk, values for $\kappa$
range from $10^{-3}$ to $10^{-1}$ cm$^2$/g (\citealt{dalessioetal01},
their Figure 1), depending on how large grains have grown; actual
values would be lower if dust were depleted relative to gas as a result
of radial drift (e.g., \citealt{andrewsetal12}).  If we adopt $\kappa
\sim 10^{-2}$ cm$^2$/g, then the optical depth of our example disk would be
$\tau = \Sigma \kappa \sim 0.1$ and the cooling time would be $10^3 (\Sigma
/ 10\, {\rm g}\, {\rm cm}^{-2}) (10\, {\rm K} / T)^3 (0.1/\tau)$ yr,
about six times longer than the local dynamical time $\Omega^{-1}
\sim 160$ yr. 
Such a disk falls squarely within the domain explored by
our optically-thin cooling simulations: for $R=100$ AU and $T=10$ K,
our $b=500$ run corresponds to an effective cooling time
of $9.64 \Omega^{-1}$ and $\kappa \sim 6 \times 10^{-3}$ cm$^{2}$/g.

Recently the long-term stability of gravito-turbulent disks has been
questioned \citep{Paardekooper2012,Hopkins2013b}. The standard
criteria for fragmentation are $\Omega\tcool \lesssim 3$ and $Q
\lesssim 1$ (e.g., \citealt{gammie01}), but a more complete assessment
of stability should account for statistical fluctuations which can
generate Toomre-unstable overdensities over long enough time intervals
(\citealt{Hopkins2013b}). We did not focus in this paper on the
dynamics of collapse. In nearly all of the parameter space that we explored
($\Omega\tcool \geq 4$, $\gamma = 5/3$, $t \sim
100$--$500\Omega^{-1}$), our 3D simulations did not show collapse. 
There were only two simulations which resulted in fragmentation: constant 
cooling with $\beta=3$ and optically-thin cooling with $b=100$ (effective
cooling times $\lesssim 3\Omega^{-1}$). Our results confirm that the standard
collapse criterion $\Omega\tcool \lesssim 3$ applies over timescales
up to dozens of orbits.

Some next steps for this work include treating optically thick disks;
accounting for stellar irradiation; and resolving radial
structure, perhaps with global simulations that
can track non-local energy and angular momentum transport.
Studies like those by Forgan et al.~(\citeyear{Forgan2012}), \citet{kratter11}, \citet{MB2010}, \citet{FR2010}, \citet{Rafikov2009}, 
\citet{LR2005}, and \citet{LR2004}
 push on these fronts, but lack the vertical
resolution that our local, optically thin simulations enjoy. 
We suspect, however, that the inclusion of additional physics will not
change our finding that gravito-turbulent velocities change by less
than a factor of 2 with vertical height. Optically thick disks will
have stronger vertical variations in cooling time than optically thin
disks, but the only cooling time that matters for determining the
overall vigor of gravito-turbulence should be that averaged over the
first scale height of the disk; only here is material actually dense
enough to be gravitationally unstable.  Whether the disk is optically
thick or thin has no bearing on the fact that material at a given
radius and height can be strongly perturbed by density fluctuations at
the midplane, either at the same radius or elsewhere.  It is the
long-range connectivity enabled by self-gravity that renders turbulent
motions the same at altitude as at depth.

\acknowledgments We thank Meredith Hughes for motivating discussions,
and Yan-Fei Jiang and Chang-Goo Kim for confirming the boundary
problem with \texttt{Athena}'s shearing-box Poisson solver.  Phil
Armitage, Xue-Ning Bai, Duncan Forgan, Charles Gammie, Yan-Fei Jiang,
Sijme-Jan Paardekooper, Jake Simon, and Jim Stone provided helpful and
encouraging comments on a draft manuscript.  We also thank an
    anonymous referee for a careful and thorough report.  Financial
support was provided by a NASA Origins grant.  Resources supporting
this work were provided by the NASA High-End Computing (HEC) Program
through the NASA Advanced Supercomputing (NAS) Division at Ames
Research Center.

\clearpage

\bibliographystyle{apj}
\bibliography{shear}

\end{document}